# Chemistry, Climate, and Transmission Spectra of TRAPPIST-1 e Explored with a Multimodel Sparse Sampled Ensemble

Eric T. Wolf[1,2,3,4], Edward W. Schwieterman[2,3,5], Jacob Haqq-Misra[2,3], Thomas J. Fauchez[3,4,6,7], Sandra T. Bastelberger[7], Michaela Leung[3,5], Sarah Peacock[3,6,8], Geronimo L. Villanueva[6], and Ravi K. Kopparapu[3,6]
[1] Laboratory for Atmospheric and Space Physics, University of Colorado Boulder, Boulder, CO 80303, USA; eric.wolf@colorado.edu
[2] Blue Marble Space Institute of Science, Seattle, WA 98104, USA
[3] Consortium on Habitability and Atmospheres of M-dwarf Planets (CHAMPs), Laurel, MD 20723, USA
[4] Sellers Exoplanet Environment Collaboration (SEEC), NASA Goddard Space Flight Center, Greenbelt, MD 20771, USA
[5] Department of Earth and Planetary Sciences, University of California, Riverside, CA 92521, USA
[6] NASA Goddard Space Flight Center, Greenbelt, MD 20771, USA
[7] Integrated Space Science and Technology Institute, Department of Physics, American University, Washington, DC 20016, USA
[8] University of Maryland, Baltimore County, Baltimore, MD 21250, USA

## Abstract

TRAPPIST-1 e is one of a few habitable zone exoplanets that is amenable to characterization in the near term. In this study our motivations are both scientific and technical. Our technical goal is to establish a multimodel sparse sampled ensemble approach for coherently exploring large unconstrained parameter spaces typical in exoplanet science. Our science goal is to determine relationships that connect observations to the underlying climate across a large parameter space of atmospheric compositions for TRAPPIST-1 e. We consider atmospheric compositions of $N_2$, $CO_2$, $CH_4$, and $H_2O$, with water clouds and photochemical hazes. We use a 1D photochemical model, a 3D climate model, and a transmission spectral model, filtered through a quasi−Monte Carlo sparse sampling approach applied across atmospheric compositions. While clouds and hazes have significant effects on the transmission spectra, $CO_2$ and $CH_4$ can be potentially detected in $\leqslant$10 transits for certain compositional and climate states. Colder climates have better prospects for characterization, due to being relatively dry and having fewer clouds, permitting transmission observations to probe more deeply into their atmospheres. $CH_4$ volume mixing ratios of $\geqslant 10^{-3}$ trigger strong antigreenhouse cooling, where near-IR absorption simultaneously creates an inversion in the stratosphere and reduces the stellar radiation reaching the planet surface. In such cases, interpreting the disk-averaged emission and albedo at face value can yield misleading conclusions, as here low albedo and high thermal emission are associated with cold planets. Future work will use our sparse sampling approach to explore broader parameter spaces and other observationally amenable exoplanets.

## 1. Introduction

Astronomy is historically a statistical science, where large numbers of photons are collected from many targets and then interpreted by comparison against expansive grids of theoretical results. To the contrary, climate science historically relies on the development of large, complex, and expensive numerical models, where a small number of detailed simulations can underpin multiple lines of scientific inquiry. Over the past decade exoplanet science has risen to prominence as an interdisciplinary field that has forced these two sciences of incongruous approaches to dance together. While the number of confirmed extrasolar planets continues to climb, uncertainties around their core quantities such as mass and radius persist, and information about their surface and atmospheric compositions is unknown in most cases. The enduring unknowns surrounding exoplanets leave a massive parameter space in their wake, needing exploration with theoretical models in order to fill in the gaps of knowledge. To date the computational expense of 3D climate models has restricted theoretical work to specific cases and contrived problems of interest.

In this work we develop a methodology for improving synergy between exoplanet astronomy and climate modeling, by performing a multimodel sparse sampled ensemble across a broad parameter space. The use of a sparse sampled ensemble allows us to run our expensive theoretical models across a much broader range of parameters compared to conventional gridding methods. We use well-established photochemical, climate, and transmission spectral models in order to produce high-quality, self-consistent simulations of our modeled worlds. We then synthesize our sparse grid using a technique known as kriging, which is related to Gaussian process emulation and allows for the rapid determination of values across large multidimensional parameter spaces (J. Haqq-Misra et al. 2024). Kriging and related Gaussian emulation techniques are able to produce reasonable results even for sparse grids with nonlinear behaviors, a scenario where conventional interpolation methods do poorly. In totality, our methodology allows us to reliably study large parameter spaces with costly theoretical models, while saving dramatically on computational expense. Ultimately, we aim to connect plausible near-term transmission spectral observations with our chemistry and climate modeling results, illuminating new ways of contextualizing exoplanet observations within planetary climatology.

The ultimate goal of any exoplanet observation is to tie the observed data to the underlying atmospheric state, including chemical composition, surface climate, and ultimately the potential for habitability. Interpretations of exoplanet observations rely on spectral modeling, which themselves critically rely on some underlying assumed, or preferably modeled, atmospheric state. Such efforts may benefit from the production and maintenance of libraries of exoplanet climate simulations that span large ranges of planetary parameters, atmospheric compositions, and surface properties, along with their associated spectral properties, including transmission, reflection, and thermal emission properties. Similar efforts have successful track records separately in both astronomy and climate science. For instance, large libraries of theoretical stellar models are widely useful across many aspects of astronomy and exoplanet sciences (e.g., F. Allard et al. 2003). In Earth climate science, the long-running Coupled Model Intercomparison Project series (e.g., V. Eyring et al. 2016) has created a rich database of theoretical model predictions of various modern climate scenarios using a variety of models, which is frequently used by the community. Recently, similar databases have been published for substellar gas giant objects (C. V. Morley et al. 2024) and for hot Jupiters (A. Roth et al. 2024); however, to date no such database has been published on terrestrial-sized extrasolar planets.

For exoplanet climate modeling, no such coherent database exists. Simple 1D climate models are sufficiently fast enough to produce expansive data sets of theoretical planetary atmospheres, but at the sacrifice of meaningful information content. Naturally, 1D models ignore horizontal gradients in temperature and composition, which are critical to climate and observation. Vertical motions are also parameterized through an ad hoc eddy diffusion coefficient, rather than explicitly computing vertical motions through convective fluxes and the action of the general circulation. Perhaps the most critical difficulty of lower-dimensional models is adequately predicting the locations of clouds and hazes, which can have outsized effects on exoplanet transmission spectra (T. J. Fauchez et al. 2019; T. D. Komacek et al. 2020; G. Suissa et al. 2020a). Clouds in particular are 4D phenomena, determined by horizontal and vertical gradients in temperature and constituent mixing ratios, driven by the general circulation, and subject to time variability (D. E. Sergeev et al. 2022a). 3D climate system models provide the most complete visions of theoretical exoplanetary climates, capturing the interconnected, time-marching, land−ocean−atmosphere system, in all spatial dimensions. However, such models are computationally expensive, making it difficult to create comprehensive databases via brute-force methods within reasonable computing availability and cost constraints. Notably, here we take some inspiration from A. Paradise et al. (2021), who conducted a grid of 480 3D climate simulations of generic M dwarf planets using the `ExoPlaSim` model (A. Paradise et al. 2022). `ExoPlaSim` is an intermediate-complexity model, with a coarse spatial resolution and pared-down physics capabilities to facilitate fast run times. Here we elect to use a full physics 3D climate model, requiring us to make concessions on the number of simulations we can feasibly run versus computational cost.

At present, the information known about any given exoplanet is generally limited to its basic physical parameters, such as the mass, radius, received stellar irradiation, and perhaps rotation rate if one can reasonably assume tidal locking for planets around M dwarf stars. However, asynchronous spin–orbit resonances are also possible for M dwarf systems (H. Chen et al. 2023). This leaves many fundamental unknowns with respect to the atmosphere and surface characteristics. The unconstrained possibilities mean that our modeling efforts must cover vast parameter spaces in order to provide sufficient background states against which observational data can be compared. To complicate matters, many atmosphere and surface characteristics can have nonlinear effects on the climate, and many such characteristics can also display degeneracies in their observable spectra. To provide comprehensive background atmospheric states for any given target, the number of required simulations can easily spiral beyond feasibility. Here, in an attempt to reconcile the demands of exoplanet astronomical observations with the computational demands of numerical climate modeling, in this study we employ a sparse sampling method called quasi–Monte Carlo (QMC) sampling (see, e.g., C. Lemieux 2009) to guide the selection of our 3D climate simulations across our parameter space of interest. Doing so allows us to reduce our computational expense by more than an order of magnitude compared to a brute-force all-grid approach. However, doing so also shifts reliance to interpolation or emulation methods to fill in the gaps.

The use of sparse gridding has a long history in climate science and more recently has been applied to exoplanet climate science. Several groups have pioneered work in applying sparse and dynamic gridding approaches for spanning large parameter spaces of exoplanet climate simulations. F. He et al. (2022) conducted a widely spaced but regular grid, to explore obliquity and eccentricity effects on exo-Earth climates. An additional layer of simulations were conducted that target regions of high sensitivity and sharp transitions. A. D. Adams et al. (2025) and N. Y. Kiang et al. (2021) both employed Latin Hypercube methods across a variety of parameters for exoplanets. A. D. Adams et al. (2025) focused on Earth-similar planets while changing orbital properties, and N. Y. Kiang et al. (2021) focused on land planets across a 10-parameter grid space. While our specific target of interest, our suite of modeling tools, and our methodology of sparse sampling are different, these works, as well as many conversations with the various parties involved, served as inspiration for our own work presented here.

This paper aims to achieve both technical and scientific goals. The primary technical goal is to demonstrate the viability of our approach, using QMC sparse sampling in tandem with multiple models, and then successfully interpolate results across our sparse grid using the kriging technique. Through this we hope to demonstrate that reliable and useful scientific information can be inferred from our technique, in the application of chemistry and climate models, in the calculation of synthetic observational predictions, and in the synthesis of all parts to develop a coherent picture of exoplanetary atmospheres across a broad parameter space. For this first study, we establish our method within a tractable parameter space, in preparation for future work by our group intending to explore much larger and more general parameter spaces. From a science standpoint, we choose to focus on modeling TRAPPIST-1 e, as it is already well studied computationally (E. T. Wolf 2017; A. P. Lincowski et al. 2018; M. Turbet et al. 2018, 2022; T. J. Fauchez et al. 2019, 2022; J. Lustig-Yaeger et al. 2019; E. May et al. 2021; D. E. Sergeev et al. 2022a; V. S. Meadows et al. 2023), while the TRAPPIST-1 system remains an object of intense focus of observation (M. Gillon et al. 2017; J. De Wit et al. 2018; L. Delrez et al. 2018; E. Ducrot et al. 2018; E. Agol et al. 2021; O. Lim et al. 2023; A. P. Lincowski et al. 2023; S. Zieba et al. 2023). Importantly, the TRAPPIST-1 system provides the most realistic opportunity to characterize terrestrial planet atmospheres within a habitable zone in the coming decade, and TRAPPIST-1 e is the most likely habitable planet candidate in the system. Thus, TRAPPIST-1 e makes for the most logical test object for our present study.

**Table 1**
Fixed Stellar and Planetary Parameters

| Parameter | Value |
|---|---|
| Stellar Model | Linear average of 1A, 2A, 2B from S. Peacock et al. (2019) |
| $T_{\rm eff}$ | 2559 K |
| $M_\star$ | 0.08 $M_\odot$ |
| $R_\star$ | 0.117 $R_\odot$ |
| Fe/H | 0.04 |
| Planet | TRAPPIST-1 e |
| Instellation | 900 W $m^{-2}$ |
| Rotation period | 6.1 days |
| Orbital period | 6.1 days |
| Planet mass | 0.772 $M_\oplus$ |
| Planet radius | 0.910 $R_\oplus$ |
| Planet density | 1.024 $\rho_\oplus$ |
| Gravity | 0.930 $g_\oplus$ |
| Surface conditions | Aquaplanet |

## 2. Methods

We conduct an ensemble of simulations using a suite of state-of-the-art models to explore the chemistry, climate, and transmission spectra of TRAPPIST-1 e. The basic parameters used for TRAPPIST-1 e and its host star are shown in Table 1 and are identical to those used in the TRAPPIST Habitable Atmospheres Intercomparison project (T. J. Fauchez et al. 2020), following from those determined by S. L. Grimm et al. (2018). In this work, we limit the atmospheric compositions studied to Archean Earth–like, having plentiful $CO_2$ and $CH_4$ in a 1-bar atmosphere with $N_2$ as the background gas, and a global ocean providing a source of $H_2O$. Water (in vapor, liquid, and ice phases) and photochemical hazes are allowed to freely evolve. While the atmospheric compositions considered here are only a subset of what is possible (M. Turbet et al. 2020), limiting our parameter space allows us to develop and demonstrate our sparse sampling methods within a manageable ensemble size. In the following subsections we describe the various models used and their relationship to each other.

### 2.1. Photochemical Modeling

First, we conduct 1D photochemical calculations using the `Atmos` model (G. Arney et al. 2016), which is freely available on Github.[9] `Atmos` is a 1D photochemical model for terrestrial planet atmospheres with heritage tracing back to the Kasting group model (J. F. Kasting et al. 1979; J. F. Kasting & J. C. G. Walker 1980; K. J. Zahnle 1986). `Atmos` has been widely used for a variety of planetary problems, including to simulate the chemistry of Archean Earth–like atmospheric compositions relevant both for the history of Earth (J. D. Haqq-Misra et al. 2008; G. Arney et al. 2016; C. E. Harman et al. 2018) and for exoplanets (G. N. Arney et al. 2017; V. S. Meadows et al. 2018; T. J. Fauchez et al. 2019), and thus is an appropriate tool for our current study. Our version of `Atmos` includes the reaction rate and $H_2O$ cross-section updates recommended by S. Ranjan et al. (2020). We use a semiempirical stellar spectrum consisting of a linear average from the three TRAPPIST-1 models (1A, 2A, 2B) presented in S. Peacock et al. (2019) and made available on the STScI website[10] by the HAZMAT team (E. L. Shkolnik & T. S. Barman 2014; B. E. Miles & E. L. Shkolnik 2017; S. Peacock et al. 2020). Stellar model 1A replicates Ly$\alpha$ observations of TRAPPIST-1 from V. Bourrier et al. (2017). Stellar models 2A and 2B represent plausible ranges of UV spectral intensities from other observed stellar M8-class stars, similar to TRAPPIST-1. The interested reader can find full details of the assumptions and character of these reference spectra in S. Peacock et al. (2019). Our nominal atmosphere has a surface pressure of 1 bar. We vary $CO_2$ logarithmically between $10^{-3}$ (1000 ppmv) and 0.5 bars (50% v/v) and vary the surface $CH_4$ flux logarithmically between 0.1 and 100 Tmol $yr^{-1}$, with $N_2$ assumed to be a filler gas. We limit our study to 1-bar total atmospheres in order to maintain a tractable parameter space. However, changing the total atmospheric pressure can have meaningful effects on climate through pressure broadening, Rayleigh scattering, lapse rate, and energy transport (Y. Zhang & J. Yang 2020).

In total, 480 Atmos simulations are run across this grid. Boundary conditions are based on those explored by G. Arney et al. (2016), but with no fixed $pO_2$. Instead, $O_2$ is determined entirely by the photochemistry of other O-bearing species and O deposition into the oceans, which allows $O_2$ to attain higher or lower $pO_2$ concentrations than in G. Arney et al. (2016). Additionally, we do not fix CO mixing ratios, and we allow CO abundances to evolve dynamically according to photochemistry with surface deposition set at the maximum biotic rate consistent with an Archean-like biosphere (P. Kharecha et al. 2005). We assume an atmosphere in hydrostatic equilibrium with a 275 K surface and a 180 K stratosphere and water vapor that follows a relative humidity profile (S. Manabe & R. T. Wetherald 1967) with a surface relative humidity of 80%. NO production by lightning is included in the model (C. E. Harman et al. 2018). The remaining detailed surface boundary conditions are given in Table A1.

Hydrocarbon hazes derived from $CH_4$ photolysis are highly complex mixtures composed of a myriad of chemical compounds and involving polymers with a large number of C atoms (e.g., M. G. Trainer et al. 2012; T. Gautier et al. 2017; S. M. Hörst et al. 2018). This renders explicit modeling computationally unfeasible. A common modeling strategy is to avoid higher-order C-bearing structures entirely by assuming cutoffs for polymerization reactions (e.g., A. A. Pavlov et al. 2001; P. Lavvas et al. 2008; V. A. Krasnopolsky 2009), the products of which are directly transformed into aerosol particles and are no longer available to reactions in the gas phase. Our Archean model uses two such hydrocarbon haze production channels: a pure polyyne pathway proceeding through the polymerization of acetylene ($C_2H_2$) by the ethynyl radical $C_2H$ and, less efficiently, via the reaction of allene ($CH_2CCH_2$) and ethynyl. Therefore, in the Atmos model, intermediate hydrocarbon haze particles are generated via the reactions $C_2H + C_2H_2 \rightarrow C_4H_2 + H$ and $C_2H + CH_2CCH_2 \rightarrow C_5H_4 + H$, with $C_4H_2$ and $C_5H_4$ assumed to directly condense into haze particles (G. Arney et al. 2016). Laboratory experiments indicate that haze production in Archean-like atmospheres can proceed when $CH_4/CO_2$ exceeds ∼0.2 (M. G. Trainer et al. 2006). Past modeling simulations with Atmos argue that spectral differences in the ultraviolet radiation control haze formation, and the $CH_4/CO_2$ threshold can vary significantly for different stellar inputs (G. N. Arney et al. 2017). For instance, G. N. Arney et al. (2017) found that for an AD Leo host star optically thick hazes would not form until the $CH_4$-to-$CO_2$ ratio rises above ∼0.9. T. J. Fauchez et al. (2019) found appreciable haze formation for the TRAPPIST-1 planets for $CH_4/CO_2 = 0.2$. More specifically, the production of haze is critically dependent on the UV flux emitted by the star at wavelengths less than 200 nm. Some flux in this region is required to drive haze photochemistry; however, too much flux can result in diminished haze production (G. N. Arney et al. 2017). This underscores the importance of observing and modeling ultraviolet spectra of different stellar types. For example, D. Teal et al. (2022) show that using different UV reconstructions and measurements results in observable differences when modeling hazy Archean-like planets. Furthermore, G. Arney et al. (2016) indicated that haze formation demonstrates a sharp sensitivity to the $CH_4/CO_2$ ratio, with a step-function-like transition from negligible haze production to thick haze production. Thus, the specific haze production for a given exoplanet is highly sensitive to both the UV spectrum and the atmospheric composition.

[9] https://github.com/VirtualPlanetaryLaboratory/atmos

[10] https://archive.stsci.edu/hlsp/hazmat

### 2.2. 3D Climate Modeling

The next step in our sequence uses the self-consistent atmospheric compositions produced by Atmos to inform our ensemble of 3D climate simulations using ExoCAM. The tidally locked nature of TRAPPIST-1 e means that to fully capture the climate, clouds, hazes, and associated observables, one must address the inherently 3D effects that can dominate the features of such worlds. This includes atmospheric circulation modulated by the planetary rotation rate and the assumption of tidal locking, day−night heat redistribution, and the horizontal distribution of clouds and hazes (J. Haqq-Misra et al. 2018; M. Cohen et al. 2024). ExoCAM is an open-source 3D climate model for exoplanets based on the National Center for Atmospheric Research's Community Earth System Model (CESM) v1.2.1 and is available publicly on Github.[11] Note that ExoCAM is run with its companion radiative transfer code ExoRT.[12] The radiative transfer model is separated into its own repository in order to facilitate offline development and testing. Henceforth when we refer to ExoCAM we are implying ExoCAM linked to ExoRT as the total package. ExoCAM has been used to study a wide variety of tidally locked exoplanets (R. Kopparapu et al. 2017; J. Haqq-Misra et al. 2018; T. D. Komacek & D. S. Abbot 2019; S. Wang & J. Yang 2022; J. Xiong et al. 2022; A. H. Lobo et al. 2023), including numerous studies specifically of TRAPPIST-1 e (E. T. Wolf 2017; A. J. Rushby et al. 2020; D. E. Sergeev et al. 2020; M. Turbet et al. 2020; E. May et al. 2021; A. Hochman et al. 2022; Y. Rotman et al. 2022). In the following paragraphs we elaborate on the specific configuration of ExoCAM used in this study; however, to avoid redundancy, we refer the reader to the model description paper (E. T. Wolf et al. 2022) and references within for a full-scope rundown on ExoCAM, ExoRT, and CESM.

ExoCAM simulations are run with a horizontal grid resolution of 4° latitude by 5° longitude. Most cases are run with 51 vertical layers, ranging from 1 bar at the surface to 0.01 mbar at the model top. For the limited set of cases where haze is included in 3D, we extend the model top a touch higher with 55 vertical layers and a 0.005 mbar model top. This is done in order to better capture the haze production region. We configure the model with a 50 m deep thermodynamic slab ocean (C. M. Bitz et al. 2012). Sea ice forms when the temperatures drop below 271.15 K, implying a salinity equal to the modern Earth's ocean. While our ocean treatment is admittedly simple and misses features compared to full-depth dynamic ocean calculations (e.g., Y. Hu & J. Yang 2014), its fast equilibration time allows us to devote our computational resources to more expensive radiation and aerosol calculations and to perform a greater number of 3D simulations within the same computational budget. A discussion of the consequences and uncertainties surrounding the choice of ocean treatment is found in Section 5.2. Two channel albedos ("visible" and "near-infrared") are calculated for sea ice and snow by weighting against the TRAPPIST-1 spectra and are 0.682 and 0.134 for sea ice and 0.978 and 0.418 for fresh snow cover, respectively (A. J. Rushby et al. 2020). Ocean albedos are taken as 0.06 at all wavelengths. Our configuration assumes the cloud and convection schemes of P. Rasch & J. Kristjánsson (1998) with an improved deep convection solver (R. P. Brent 1973) and the finite-volume dynamical core (S.-J. Lin & R. B. Rood 1996). In our ExoCAM configuration, the model reaches equilibrium climate states in ∼100 Earth years. However, in frozen cases, very slow cooling can persist on the night hemisphere for numerous centuries as surface snow and ice build up indefinitely, releasing latent heat in the process. The nature of geologic timescale formation of glacier ice on tidally locked planets is beyond the scope of the model and of this work.

Our radiation model, ExoRT, is built around the standard (O. B. Toon et al. 1989) two-stream solver to calculate the radiative transfer across 68 spectral bins. Both the shortwave (i.e., stellar) and longwave (i.e., planetary) energy streams share a common spectral grid and a common set of absorption coefficients. The spectral range of integration is then set independently for both energy streams. ExoRT has recently been updated to optimize the shortwave and longwave bandpasses automatically based on the input stellar spectra, assumptions on the maximum planetary temperatures, and a truncation tolerance (currently set at 0.1%). A calculation of shortwave and longwave energy across all 68 spectral bins for each radiation stream is unnecessary since some spectral intervals are effectively devoid of energy. Here we optimize around the TRAPPIST-1 stellar energy distribution of A. P. Lincowski et al. (2018) in the shortwave and for 400 K maximum temperatures in the long wave. The shortwave stream is calculated over 53 intervals from 22.7273 to 0.45 $\mu$m, and the longwave stream is calculated over 36 intervals from infinity to 2.65947 $\mu$m. For the shortwave, the remaining bands are renormalized to the total stellar irradiance set by the user, thus removing any energy loss from spectral truncation. In the long wave, truncation on the fly is not possible without introducing inconsistencies between the surface energy balance calculations and the atmosphere; thus, we tune for a temperature higher than experienced in our simulations. Note that the TRAPPIST-1 spectra are very red, so our optimization scheme skews toward including longer wavelengths, while omitting some short-wavelength bins that have little energy.

[11] https://github.com/storyofthewolf/ExoCAM
[12] https://github.com/storyofthewolf/ExoRT

In this study, our atmospheres contain $H_2O$, $CO_2$, $CH_4$, and $N_2$ only. Correlated-*k* distributions are calculated for $H_2O$, $CO_2$, and $CH_4$ using the HITRAN 2016 spectroscopic database (I. E. Gordon et al. 2017) in conjunction with the `HELIOS-K` accelerated *k*-coefficient sorter (S. L. Grimm & K. Heng 2015). Self and foreign water vapor continuum coefficients are included using MT_CKD version 3.2. Relevant collision-induced absorption pairs of $CO_2$ and $N_2$ are also included, along with a standard implementation of Rayleigh scattering contributions from all species (I. M. Vardavas & J. H. Carver 1984). Cloud and ice particles are treated as Mie scattering particles. `ExoRT` uses the equivalent extinction method to handle overlapping gases (D. S. Amundsen et al. 2017), with a dynamic selection of major versus minor gas species continuously recalculated in the model. In the equivalent extinction method, the primary gas absorber in each spectral interval is treated with an 8 G point *k*-distribution, while all additional species that have absorption in the band are treated as gray absorbers using their band-averaged absorption value.

Atmospheric compositions in `ExoCAM` assume constant mixing ratio profiles for $CO_2$, $CH_4$, and $N_2$ informed by `Atmos`. Although photochemical calculations indicate that the mixing ratios of $CO_2$ and $CH_4$ begin to slightly decrease above 0.1 mbar owing to photochemical destruction winning over vertical mixing, the effects on the radiation field are negligible. For atmospheric compositions where `Atmos` predicts optically thick photochemical hazes to form, we use the Community Aerosol and Radiation Module for Atmospheres (`CARMA`; R. P. Turco et al. 1979; O. B. Toon et al. 1988; C. G. Bardeen et al. 2008), which is fully coupled to `ExoCAM`. `CARMA` is a flexible full physics cloud and aerosol model that can be configured to treat any manner of cloud and aerosol problem, although it is computationally expensive. `CARMA` has already been used to simulate hazes in 3D for early Earth (E. T. Wolf & O. B. Toon 2010) and Titan (E. J. L. Larson et al. 2014, 2015). `Atmos` calculations provide zenith angle, altitude, and atmospheric-composition-dependent haze production rate profiles that are used to source haze monomer particle creation in `ExoCAM-CARMA`, which then freely evolve in the 3D model. We assume fractal aggregate hazes following directly from E. T. Wolf & O. B. Toon (2010). Note that a similar haze implementation was ported to `Atmos` and was used in G. Arney et al. (2016), G. N. Arney et al. (2017), T. J. Fauchez et al. (2019), and elsewhere.

We acknowledge that laboratory work on measuring haze refractive indices of various compositions is a vibrant and critical area of research (e.g., S. M. Hörst et al. 2018; C. He et al. 2023), but here we use the well-trodden refractive indices of B. N. Khare et al. (1984). The B. N. Khare et al. (1984) refractive indices have been dominantly used in past modeling studies of photochemical hazes (e.g., J. D. Haqq-Misra et al. 2008; G. Arney et al. 2016; G. N. Arney et al. 2017; E. J. L. Larson et al. 2014, 2015; T. J. Fauchez et al. 2019; P. Gao et al. 2023; M. Cohen et al. 2024) and thus make for a natural starting place. These were produced under laboratory conditions containing 90% $N_2$ and 10% $CH_4$, which are similar to our hazy atmospheres. Recent laboratory measurements of haze refractive indices produced in more early Earth-like environments with $CO_2$ and only few percent $CH_4$ indicate a reasonable agreement with the B. N. Khare et al. (1984) values with respect to overall spectral shape (see Figure 12 in T. Drant et al. 2024). To produce optical constants from the refractive indices, we use the fractal mean field theory to produce optical constants (R. Botet et al. 1997), which was converted to FORTRAN 90 and preserved on Github.[13]

Simulations are started from a warm initial condition with a 300 K isothermal temperature field, no water and no clouds in the atmosphere, and zero initial winds. This is identical to the protocol used in the Trappist Habitable Atmospheres Intercomparison (THAI) project (T. J. Fauchez et al. 2020). For hazy simulations that utilize `CARMA`, we first spin up the simulations without haze and then add hazes to the already equilibrated haze-free results, bootstrapping these simulations to save on computational cost. Note that D. E. Sergeev et al. (2022b) found that TRAPPIST-1 e could exhibit a bistability of dynamical regimes with either single-jet and double-jet states possible depending on the fine details of the initial conditions. J. H. Checlair et al. (2019) found that terrestrial planets around M dwarf stars are unlikely to exhibit snowball hysteresis owing to the red incident stellar spectra, reduced ice albedos in the near-infrared, and infinite length of solar days. Thus, our results would unlikely be changed by cold start conditions. However, M. Turbet et al. (2021) found that, if initialized with an excessively hot steam atmosphere, terrestrial planets may never cool sufficiently to allow surface liquid water. Parsing the effects of variances in initial conditions is beyond the scope of this work. We adhere to the established THAI protocol for initializing our 3D simulations.

### *2.3. Sparse Sampling and Grid Interpolation*

Each `ExoCAM` simulation as configured here takes a minimum ∼20,000 core-hours to complete. `CARMA`-enabled simulations cost nearly a factor of 3 more than haze-free simulations. Thus, we bootstrap hazy simulations from haze-free counterparts to reduce the computational burden in reaching climatic equilibrium. Even so, the expense of running the 3D climate model makes it difficult to simulate large ensembles of unique simulations. Recall that our first phase of 1D photochemical modeling contains 480 unique cases. Simulating all 480 unique cases with `ExoCAM` would require a prohibitive amount of computing time, consuming our entire 5 yr project's computational budget. Thus, we must be selective in how and when we deploy our 3D climate model. Here we use a QMC sparse sampling technique to guide the application `ExoCAM` across our 2D parameter space of $CO_2$ and $CH_4$. QMC selection is designed to provide representative samples across the complete multidimensional parameter space, without leaving spurious gaps that could arise from using conventional pseudorandom number generators. Our particular implementation of QMC uses a low-discrepancy sequence (i.e., one with good space-filling properties) known as a scrambled Sobol sequence (S. Joe & F. Y. Kuo 2003, 2008), which carries the additional requirement that each sequence of uniformly distributed points in the parameter space be equal to $2^n$, where $n$ is any integer. Further examples of this application of QMC are described in J. Haqq-Misra et al. (2022). In total we select 32 of the original 480 atmospheric compositions to simulate with ExoCAM.

[13] https://github.com/storyofthewolf/fractal_optics_coreshell

After completing our 32 3D climate simulations, we must then synthesize our sparse grid of climate and observable quantities into meaningful information. Here we use a technique known as "kriging" to synthesize our sparse sampled grid. Kriging was originally developed as a method to estimate spatial distributions of precious metal reserves based on sparse sampled drill cores. It has since been applied to a wide range of problems, including climate science (D. Drignei 2009; S. Garrigues et al. 2021), astrophysics (M. Tremmel et al. 2017), and exoplanets (J. Haqq-Misra et al. 2024). Kriging is most closely related to Gaussian process emulation, with the former having its origins in geostatistics and the latter more widely used in computer science and machine learning. The language and mathematics used to describe these two methods differ, but in general Gaussian process emulation can be considered as " the expansion of the kriging method to the probabilistic framework under the Gaussian assumption" (M. Marinescu 2024); in other words, the two methods share many similarities but are not always identical. However, in situ ations where the mean is unknown (as in this study), the two methods can provide identical results. It is worth noting that the "ordinary kriging" approach used in this study provides exact interpolation at each point in a sparse sample, whereas many applications of Gaussian process emulation typically use larger sets of data to construct surrogate models that are accurate to within an acceptable error threshold. The use of kriging in this study will provide a way to visualize and interpolate the results of the 32 cases, which could also be developed into a kriging-based surrogate model in future work.

### 2.4. Transmission Spectral Modeling

For the last step in our atmospheric modeling process, we take the results from our 32 3D climate simulations with ExoCAM and import them to the Global Emission Spectra (GlobES) application of the Planetary Spectrum Generator[14] (PSG; G. L. Villanueva et al. 2018, 2022; T. J. Fauchez et al. 2025) to calculate theoretical transmission spectra. PSG is an online radiative transfer code combining the latest radiative transfer methods and spectroscopic parameterizations, and it includes a realistic treatment of multiple scattering in layer-by-layer spherical geometry. Exoplanet atmospheres, specifically for tidally locked exoplanets, as assumed for TRAPPIST-1 e, can be highly heterogeneous. ExoCAM-generated 3D atmospheric fields contain temperature; pressure; volume mixing ratios of molecular species $N_2$, $CO_2$, $CH_4$, and $H_2O$; mass mixing ratios and particle sizes for liquid and ice water clouds; and fractal aggregate hazes where appropriate. GlobES is used to ingest all the ExoCAM netCDF outputs and produce planetary spectra, for any kind of atmosphere and for any observing geometry and instrument mode. GlobES leverages the well-tested PSG radiative transfer suite to accurately compute fluxes, transmittances, reflectances, and emissivities for a wide range of planetary objects. Multiple scattering from atmospheric aerosols is computed using the discrete ordinates method, which assumes a plane-parallel atmosphere in which the radiation field is approximated by a discrete number of streams angularly distributed. Legendre polynomials are used to represent the angular dependence of the aerosol scattering phase function. For each available aerosol type in PSG (e.g., S. T. Massie & M. Hervig 2013), look-up tables of Legendre expansion coefficients are precomputed using an assumed particle size distribution. For this project, we created a new aerosol input file specific for fractal hazes, with B. N. Khare et al. (1984) refractive indices, following the fractal aggregate prescription described in E. T. Wolf & O. B. Toon (2010) and using the fractal mean field theory (R. Botet et al. 1997) to derive the extinction and scattering coefficients and asymmetry parameter. PSG contains billions of spectral lines of over 1000 species from several constantly updated spectroscopic repositories (e.g., HITRAN, HITEMP, ExoMol, JPL, CDMS, GSFC-Fluor). In this work, we used the HITRAN 2020 database (I. E. Gordon et al. 2022), which is complemented by UV/optical data from the MPI database (H. Keller-Rudek et al. 2013).

In addition to computing transmission spectra, we also calculated the number of transits required to obtain a $5\sigma$ detection of the gas components of our atmosphere following from past works (J. Lustig-Yaeger et al. 2019; T. J. Fauchez et al. 2020, 2022), where $N_{\text{transits}}^{5\sigma} = (5/\text{SNR})^2$ and SNR is the signal-to-noise ratio of the gas of interest integrated over the full spectral range. SNR is calculated as the quadratic sum, for each spectral bin, of the difference between the spectrum with and without the gas of interest, divided by the noise (T. J. Fauchez et al. 2022). After compilation of the transmission spectral results, we also interpolate these observable quantities across the parameter space from our grid of models using kriging, as discussed in Section 4.

## 3. Results

### 3.1. 1D Photochemical Modeling

We use the photochemical model Atmos to compute the steady-state atmospheric $CH_4$ mixing ratios and haze production rates as a function of $CO_2$ abundance and $CH_4$ surface fluxes for an anoxic Archean-like TRAPPIST-1 e. In total 480 simulations were conducted with $CO_2$ ranging from 1000 ppmv to 0.5 bars and $CH_4$ surface flux ranging from 0.1 to 100 Tmol per Earth year. Our results are summarized in Figure 1. The left panel shows the resulting steady-state atmospheric $CH_4$ volume mixing ratios, and the right panel shows the photochemical haze steady-state column density calculated by Atmos. Naturally, higher $CH_4$ surface fluxes are associated with higher sustained atmospheric $CH_4$ concentrations. There are two primary sinks for $CH_4$ in anoxic atmospheres: reaction with OH sourced from $H_2O$ photolysis, which is most important at low $CH_4$ concentrations (or fluxes), and direct photolysis of $CH_4$ by far-UV photons (e.g., Ly$\alpha$ and the far-UV continuum), which is most important at high $CH_4$ mixing ratios and fluxes (W. Broussard et al. 2024). In general, $CH_4$ concentrations are larger when higher $CO_2$ concentrations are also present (given the same methane flux) owing to overlying $CO_2$, which shields both underlying $H_2O$ and $CH_4$ from photolysis. This relationship is not perfectly smooth because $CO_2$ photolysis also liberates O atoms that can be an additional, tertiary sink for $CH_4$ and other hydrocarbons (H.-b. Chen et al. 2005; G. N. Arney et al. 2017). The formation of hydrocarbon hazes also influences the predicted $CH_4$ mixing ratios, but the effect is confined to a relatively narrow slice of our parameter space where optically thick hazes are permitted to form. Note that while other species concentrations, for instance, CO and $C_2H_6$, are produced by `Atmos` simulations, only $N_2$, $CO_2$, $CH_4$, and hazes are accepted as inputs into our 3D model `ExoCAM` at the time of this study.

[14] https://psg.gsfc.nasa.gov/apps/globes.php

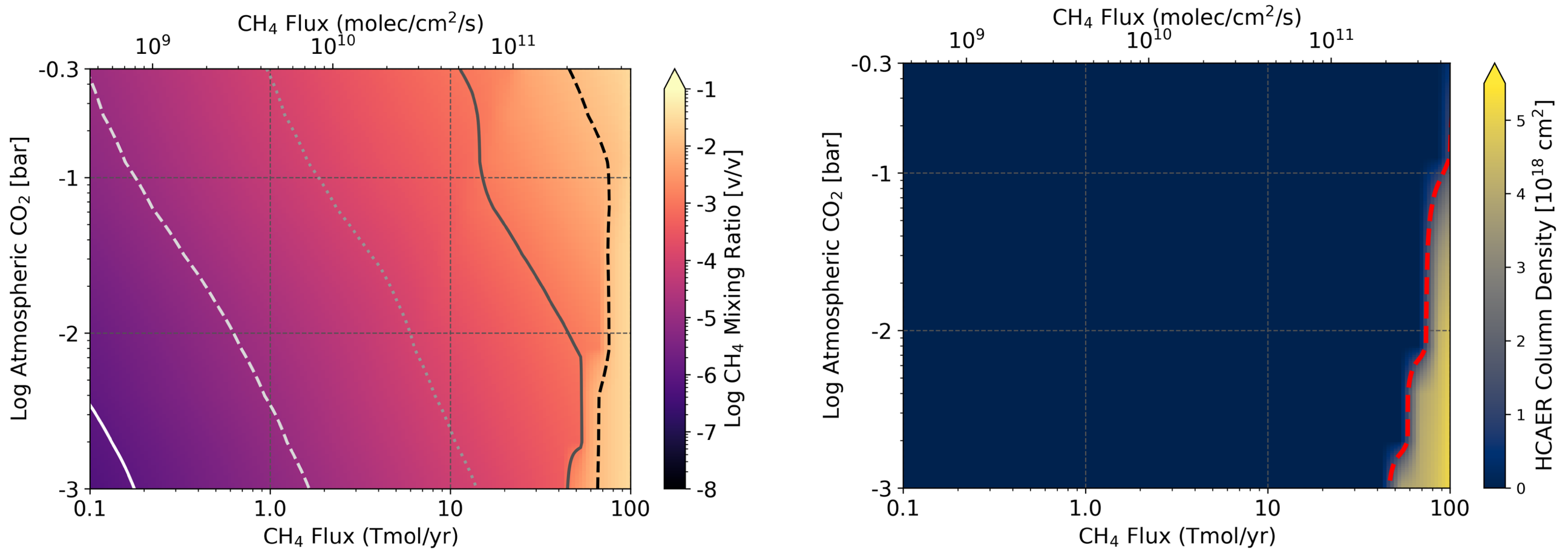

**Figure 1.** Photochemical calculations for a TRAPPIST-1 e anoxic atmosphere for a wide variety of $CO_2$ pressure logarithmically spaced between $10^{-3}$ (1000 ppmv) and 0.5 bars (50% v/v) and $CH_4$ flux logarithmically spaced between 0.1 and 100 Tmol $yr^{-1}$ (approximately between $4 \times 10^9$ molecules $cm^{-2}$ $s^{-1}$ and $5 \times 10^{11}$ molecules $cm^{-2}$ $s^{-1}$). The left panel shows surface $CH_4$ mixing ratios. The contour lines correspond to $CH_4$ mixing ratios of 1, 10, 100, 1000, and 10,000 ppm (left to right). The right panel shows the column density of hydrocarbon haze particles in $cm^2$. The dashed red line indicates the approximate $\tau = 1$ extinction contour at 550 nm. Only at the highest $CH_4$ does haze become significant. Calculations assume a 1-bar atmosphere with $N_2$ as the filling gas and include the formation of fractal hydrocarbon hazes.

The right panel of Figure 1 shows the photochemical haze column density calculated by `Atmos`. While trace amounts of haze are produced at all parts of our parameter space, only in the lower right corner of the grid do we find that an optically thick photochemical haze layer can be formed. This region of the grid features high $CH_4$-to-$CO_2$ ratios, which are canonically associated with haze formation. The approximate $\tau = 1$ haze extinction optical depth at 550 nm is indicated in the figure, though we note that the extinction optical depth uniquely attributable to hydrocarbon haze is strongly dependent on wavelength and somewhat dependent on the other atmospheric constituents (e.g., $CH_4$ and $CO_2$) present in the atmosphere. We note that while the *y*-axis in Figure 1 shows $CH_4$ in terms of surface flux, when considering the $CH_4$-to-$CO_2$ ratio with respect to haze formation, one must consider the steady-state volume mixing ratios of each gas. In our simulations, we find that significant hazes form only when $CH_4/CO_2 \geqslant 0.2$, with an exceedingly sharp transition between thick hazes and vanishing hazes. Hydrocarbon hazes also produce a shielding effect that can allow for greater buildup of $CH_4$ and thus yield a positive feedback on haze formation rates. However, because higher-order hydrocarbons, including haze precursors, are also a sink for $CH_4$, the relationship between $CH_4$ flux and haze is complex.

Our results regarding hazes are not dissimilar to past works. G. N. Arney et al. (2017) found for AD Leo, using the spectrum of A. Segura et al. (2005), that $CH_4/CO_2 = 0.9$ was required before significant hazes could form. T. J. Fauchez et al. (2019) found that significant hazes were possible for the TRAPPIST-1 planets with $CH_4/CO_2 = 0.2$, in general agreement with our current work. However, T. J. Fauchez et al. (2019) did not explore other $CH_4/CO_2$ ratios to probe the threshold value. The body of work on haze production indicates that the specific structure of near-, mid-, and far-UV stellar fluxes governs the rate of haze formation vis-à-vis the atmospheric composition. G. N. Arney et al. (2017) identified the 140–160 nm region, coincident with the peak of the $CO_2$ UV cross section, as perhaps the leading predictor of haze formation, with high incident fluxes in this spectral region leading to lower haze production rates owing to $CO_2$ photolysis and O atom liberation as noted above. To complicate matters, the ultraviolet emission from stars depends on its activity level (A. P. Lincowski et al. 2018; S. Peacock et al. 2019), which is not easily predicted from basic stellar properties and could drive time-dependent atmospheric chemistry (H. Chen et al. 2021). Literature disagreements in the modeled steady-state TRAPPIST-1 UV spectrum have been demonstrated to substantially impact $O_3$ columns in $O_2$-rich atmospheres (G. J. Cooke et al. 2023) and could similarly impact our results here. For our present study, we consider the steady-state haze production rates as a function of zenith angle, which are input into our 3D climate simulations in Section 3.3.

### *3.2. Quasi–Monte Carlo Sparse Sampling*

Our photochemical grid simulations are not predictive of the atmospheric structure, cloudiness, or the climate, only the chemical composition and haze production rates. To produce a complete picture of the atmosphere, relevant for determining the climate and for determining potential observable measurements, we must augment our `Atmos` calculations with a more complex model, using the predicted chemical compositions and haze production rates as input parameters. 3D climate models can provide a self-consistent picture of exoplanet atmospheres, albeit at a steep computational cost. The photochemical grid of simulations described in Section 3.1 contains 480 simulations, which, while not impossible to duplicate with our 3D climate model, would be extraordinarily burdensome in terms of computational expense, wall-clock time, storage space, and person hours to manage such an endeavor. Therefore, here we use a QMC sparse sampling scheme following the approach described in Section 2.3 to subsample the 480-point photochemical grid for running a limited but representative set of cases with the 3D climate model.

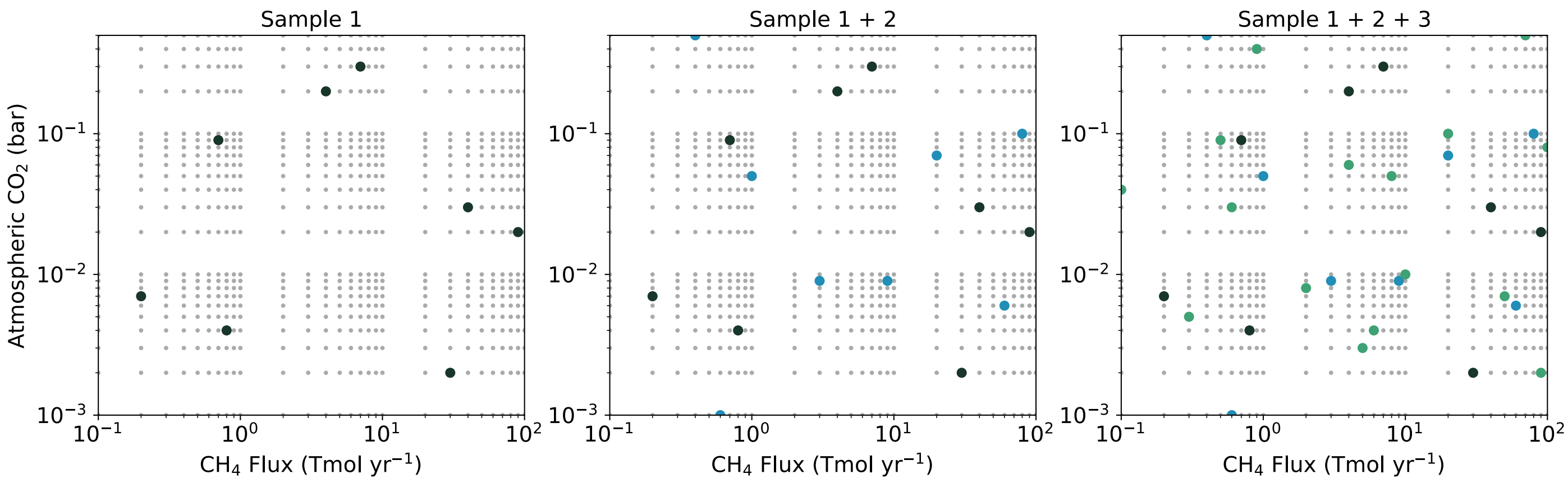

**Figure 2.** QMC selection of points from the photochemical model parameter space. The total sequence of 32 points is separated into three samples: the first 8 cases (left), the next 8 cases (middle), and the remaining 16 cases (right).

We select a total of 32 points from the photochemical calculations, separated into three samples of 8, 8, and 16 cases, as shown in the progression of panels in Figure 2. This QMC selection uses a scrambled Sobol sequence as the specific algorithm, with the seed value fixed at 3,850,109 for reproducibility. The Sobol sequence is chosen because it enables additional samples to be drawn to extend the sequence as needed and still retain the same space-filling properties across the parameter space. Figure 2 shows the sampling of the parameter space in terms of $CO_2$ partial pressure versus the $CH_4$ surface flux, the same as Figure 1 for the complete set of `Atmos` simulations. We have chosen to explicitly show three sets of points selected with the Sobol sequence in order to provide a means of quantifying the improvement (and associated errors) that occurs from increasing the number of sampled points. This will be discussed further in Section 4. Table 2 lists the $CO_2$ and $CH_4$ volume mixing ratios for each of the 32 sampled cases. These values are input in `ExoCAM` to define the atmospheric composition for 3D climate modeling in Section 3.3. Note that Table 2 also contains the $CH_4/CO_2$ ratio, which is a leading indicator for haze formation as discussed in Section 3.1. Our photochemical results reveal that 3 of our 32 sampled cases would form optically thick hazes: cases 7, 23, and 26, which have $CH_4/CO_2$ ratios of 1.2, 12, and 0.3875, respectively. For these three cases we include hazes in our 3D climate simulations using `CARMA`. From `Atmos` calculations we determined the haze production rate of these cases to be $\sim 10^{14}$ g per Earth year. See Section 3.3 for a full discussion of our hazy planet results. Note that three additional cases fall into a thin transition region between thick and vanishingly thin hazes: cases 10, 15, and 31, which have $CH_4/CO_2$ ratios of 0.096, 0.18, and 0.14, respectively. However, their haze production rates are a factor of $10^4$–$10^6$ less than for our three thick haze cases. This behavior is in agreement with G. N. Arney et al. (2017), who found a sharp transition between negligible and thick hazes as a function of $CH_4$ mixing ratios. We performed a sensitivity test, running case 10 with `ExoCAM-CARMA` and a haze production rate of $\sim 10^{10}$ g per Earth year, and we found there to be no discernible effect on climate and only a negligible effect on the resulting transmission spectra, virtually indistinguishable from haze-free in observations. Future studies could examine precisely at which $CH_4/CO_2$ ratios and haze production rates hazes become optically important. Here our sparse grid approach prevents us from quantifying this transition more tightly.

### *3.3. 3D Climate Modeling*

We use `ExoCAM` to simulate 32 different atmospheric compositions for TRAPPIST-1 e, as a sampling of the 480-simulation grid of `Atmos` photochemical calculations shown in Figure 1, using the QMC-selected points illustrated in Figure 2. Table 2 lists the atmospheric composition and global mean climate statistics for each of the 32 `ExoCAM` simulations, including the global mean surface temperature, top-of-atmosphere albedo, sea-ice fraction, and outgoing longwave radiation (OLR). Out of the 32 total `ExoCAM` cases, 29 could be classified as surface habitable, if using the most generous of definitions where any nonzero fraction of the surface having open ocean waters constitutes a habitable world (e.g., T. Jansen et al. 2019). Of the 29 surface habitable cases, global mean surface temperatures ($T_S$) range from 244 to 295 K, while maximum surface temperatures range from 278 to 314 K. Global mean sea-ice fractions range from 0.02 in our warmest simulation to 0.92 for our coldest nonsnowball case. The existence of stable climates with sea-ice fractions greater than 0.9 indicates that there is no sharp bifurcation to snowball states. In agreement with previous studies, snowball transitions are gradual for planets around M dwarf stars owing to the low albedo of snow and sea ice in the near-infrared (A. L. Shields et al. 2014; J. H. Checlair et al. 2019) and the fixed zenith angle of incident stellar radiation on the planet. Of course, ocean dynamics and sea-ice drift, coupled with variations in continental positioning, which have not been considered here, can have significant effects on the sea-ice coverage. The consequences and significant underlying uncertainties surrounding these factors are considered in Section 5.2.

Most of our simulations would be classified as cold but habitable, with 22 out of the 29 surface habitable cases having greater than half the planet covered with sea ice and global mean temperatures at or below 274 K. Three of our `ExoCAM` cases enter a complete snowball glaciation, where the entire surface is covered by a solid layer of sea ice, with global mean

**Table 2**
Global Mean Climate Statistics for Haze-free Simulations

| Sample | Cases | $CO_2$ (vmr) | $CH_4$ (vmr) | $CH_4/CO_2$ | $T_S$ (K) | Albedo | Ice Fraction | OLR ($Wm^{-2}$) |
|---|---|---|---|---|---|---|---|---|
| Sample 1 | 1 | 0.007 | $3.3 \times 10^{-6}$ | $4.7 \times 10^{-4}$ | 249.9 | 0.22 | 0.73 | 176.9 |
| | 2 | 0.03 | $1.7 \times 10^{-3}$ | $5.7 \times 10^{-2}$ | 257.5 | 0.11 | 0.78 | 200.9 |
| | 3 | 0.002 | $3.1 \times 10^{-4}$ | 0.16 | 252.0 | 0.17 | 0.74 | 188.5 |
| | 4 | 0.2 | $3.1 \times 10^{-4}$ | $1.6 \times 10^{-3}$ | 285.8 | 0.10 | 0.20 | 203.6 |
| | 5 | 0.004 | $9.5 \times 10^{-6}$ | $2.8 \times 10^{-3}$ | 248.6 | 0.22 | 0.74 | 177.0 |
| | 6 | 0.3 | $6.0 \times 10^{-4}$ | $2.0 \times 10^{-3}$ | 294.5 | 0.09 | 0.02 | 204.1 |
| | 7[a] | 0.02 | $2.4 \times 10^{-2}$ | 1.20 | 225.4 | 0.11 | 1.00 | 201.5 |
| | 8 | 0.09 | $4.2 \times 10^{-5}$ | $4.7 \times 10^{-4}$ | 275.4 | 0.14 | 0.48 | 194.3 |
| Sample 2 | 9 | 0.001 | $8.6 \times 10^{-7}$ | $8.6 \times 10^{-4}$ | 244.4 | 0.23 | 0.76 | 173.6 |
| | 10 | 0.1 | $9.6 \times 10^{-3}$ | $9.6 \times 10^{-2}$ | 244.0 | 0.06 | 0.92 | 211.5 |
| | 11 | 0.009 | $1.4 \times 10^{-4}$ | $1.6 \times 10^{-2}$ | 258.5 | 0.16 | 0.71 | 189.6 |
| | 12 | 0.05 | $4.4 \times 10^{-5}$ | $8.8 \times 10^{-4}$ | 270.7 | 0.15 | 0.62 | 191.5 |
| | 13 | 0.009 | $4.9 \times 10^{-5}$ | $5.4 \times 10^{-3}$ | 256.2 | 0.18 | 0.71 | 185.0 |
| | 14 | 0.07 | $1.2 \times 10^{-3}$ | $1.7 \times 10^{-2}$ | 263.7 | 0.12 | 0.72 | 199.6 |
| | 15 | 0.006 | $1.1 \times 10^{-3}$ | 0.18 | 255.4 | 0.13 | 0.75 | 196.2 |
| | 16[b] | 0.5 | $4.5 \times 10^{-5}$ | $9.0 \times 10^{-5}$ | 286.0 | 0.11 | 0.15 | 199.8 |
| Sample 3 | 17[b] | 0.005 | $4.2 \times 10^{-6}$ | $8.4 \times 10^{-4}$ | 249.2 | 0.22 | 0.74 | 175.9 |
| | 18[b] | 0.5 | $1.7 \times 10^{-2}$ | $3.4 \times 10^{-2}$ | 244.3 | 0.06 | 0.88 | 211.3 |
| | 19 | 0.01 | $1.7 \times 10^{-4}$ | $1.7 \times 10^{-2}$ | 259.9 | 0.15 | 0.70 | 191.4 |
| | 20 | 0.06 | $1.8 \times 10^{-4}$ | $3.0 \times 10^{-3}$ | 275.3 | 0.13 | 0.37 | 196.3 |
| | 21 | 0.008 | $3.1 \times 10^{-5}$ | $3.9 \times 10^{-3}$ | 253.9 | 0.19 | 0.72 | 182.5 |
| | 22 | 0.05 | $3.0 \times 10^{-4}$ | $6.0 \times 10^{-3}$ | 273.3 | 0.12 | 0.48 | 198.7 |
| | 23[a,b] | 0.002 | $2.4 \times 10^{-2}$ | 12.0 | 226.1 | 0.12 | 1.00 | 199.3 |
| | 24 | 0.09 | $3.1 \times 10^{-5}$ | $3.4 \times 10^{-4}$ | 274.2 | 0.15 | 0.53 | 192.8 |
| | 25 | 0.03 | $2.1 \times 10^{-5}$ | $7.0 \times 10^{-4}$ | 262.2 | 0.18 | 0.68 | 185.9 |
| | 26[a] | 0.08 | $3.1 \times 10^{-2}$ | 0.39 | 228.0 | 0.08 | 1.00 | 207.8 |
| | 27 | 0.004 | $6.6 \times 10^{-5}$ | $1.7 \times 10^{-2}$ | 252.5 | 0.19 | 0.73 | 183.6 |
| | 28 | 0.4 | $9.3 \times 10^{-5}$ | $2.3 \times 10^{-4}$ | 287.7 | 0.10 | 0.13 | 202.5 |
| | 29 | 0.003 | $4.8 \times 10^{-5}$ | $1.6 \times 10^{-2}$ | 250.3 | 0.20 | 0.73 | 181.3 |
| | 30 | 0.1 | $1.5 \times 10^{-3}$ | $1.5 \times 10^{-2}$ | 264.5 | 0.11 | 0.72 | 200.5 |
| | 31 | 0.007 | $9.6 \times 10^{-4}$ | 0.14 | 256.7 | 0.13 | 0.74 | 196.6 |
| | 32 | 0.04 | $4.5 \times 10^{-6}$ | $1.1 \times 10^{-4}$ | 261.1 | 0.19 | 0.68 | 183.8 |

**Notes.**
[a] 3D simulation contains photochemical hazes.
[b] Denoted as a representative case.

surface temperatures below 228 K. These three cases (7, 23, 26) are characterized by low to moderate $CO_2$, high $CH_4$, and photochemical haze formation. As is discussed in detail a bit later in this section, a strong antigreenhouse effect is present owing to near-infrared absorption by $CH_4$ in the upper atmosphere, creating a stratospheric inversion and preventing stellar energy from penetrating down to the planet surface. This setup results in exceedingly cold surface temperatures. We find that the haze layer plays a lesser role in governing the overall climate, but it does play a strong role in modifying transmission spectra, as is discussed in Section 3.4.

Figure 3 summarizes the global mean surface temperature results as functions of $CO_2$ and $CH_4$ volume mixing ratios. Both panels have $T_S$ as the $y$-axis and have $CO_2$ or $CH_4$ as the $x$-axis, with the data points color-coded by the mixing ratio of the alternate gas. In plotting this way, the general organization of the ensemble begins to emerge. In the left panel we see that most of the ExoCAM simulations warm monotonically with increasing $CO_2$ following a logarithmic dependence on the $CO_2$ mixing ratios, a trend that can be inferred purely from radiative transfer considerations (D. M. Romps et al. 2022). However, for high values of $CH_4$ ($\geqslant 10^{-3}$) the pattern of monotonic warming breaks. High-$CH_4$ simulations result in significantly colder planetary temperatures regardless of the background $CO_2$ that is present. Five cases break with the expected warming trend (cases 7, 10, 18, 23, 26), all featuring high $CH_4$, which drives antigreenhouse cooling of the surface. In the right panel of Figure 3 we see that as $CH_4$ is increased from small values up to $\sim 10^{-3}$ the greenhouse effect dominates and the planet warms. However, for still higher values of $CH_4$ global mean temperatures drop sharply.

Figure 4 summarizes global mean quantities for ice fraction, top-of-atmosphere albedo, OLR, and column amounts of water vapor, liquid water cloud, and ice water cloud, plotted versus the global mean surface temperature on the $x$-axis, with data points color-coded by the $CH_4$ volume mixing ratio. Hydrological cycle quantities, including sea-ice fraction, water vapor, and liquid water clouds, are tethered smoothly to the surface temperature, regardless of greenhouse composition. Ice coverage varies inversely to the surface temperature, as would be expected. Note that a region of relative stability is observed for climates with global mean temperatures between 245 and 265 K, where little change in the ice fraction is found with planet warming. However, a sharper change toward ice melting is seen for climates with global mean temperatures $\geqslant$270 K, before the planet becomes virtually ice free by $T_S$ = 295 K. Water vapor and liquid water cloud column amounts generally increase smoothly with increasing global mean temperatures. Warmer planets promote increased surface evaporation and increased convection and allow the atmosphere to hold more water. Ice water clouds appear to have a sweet spot, with maximum amounts found when global mean temperatures are between 240 and 280 K. For lower planet temperatures, water vapor availability is curtailed, which limits clouds, while for warmer temperatures liquid water clouds dominate.

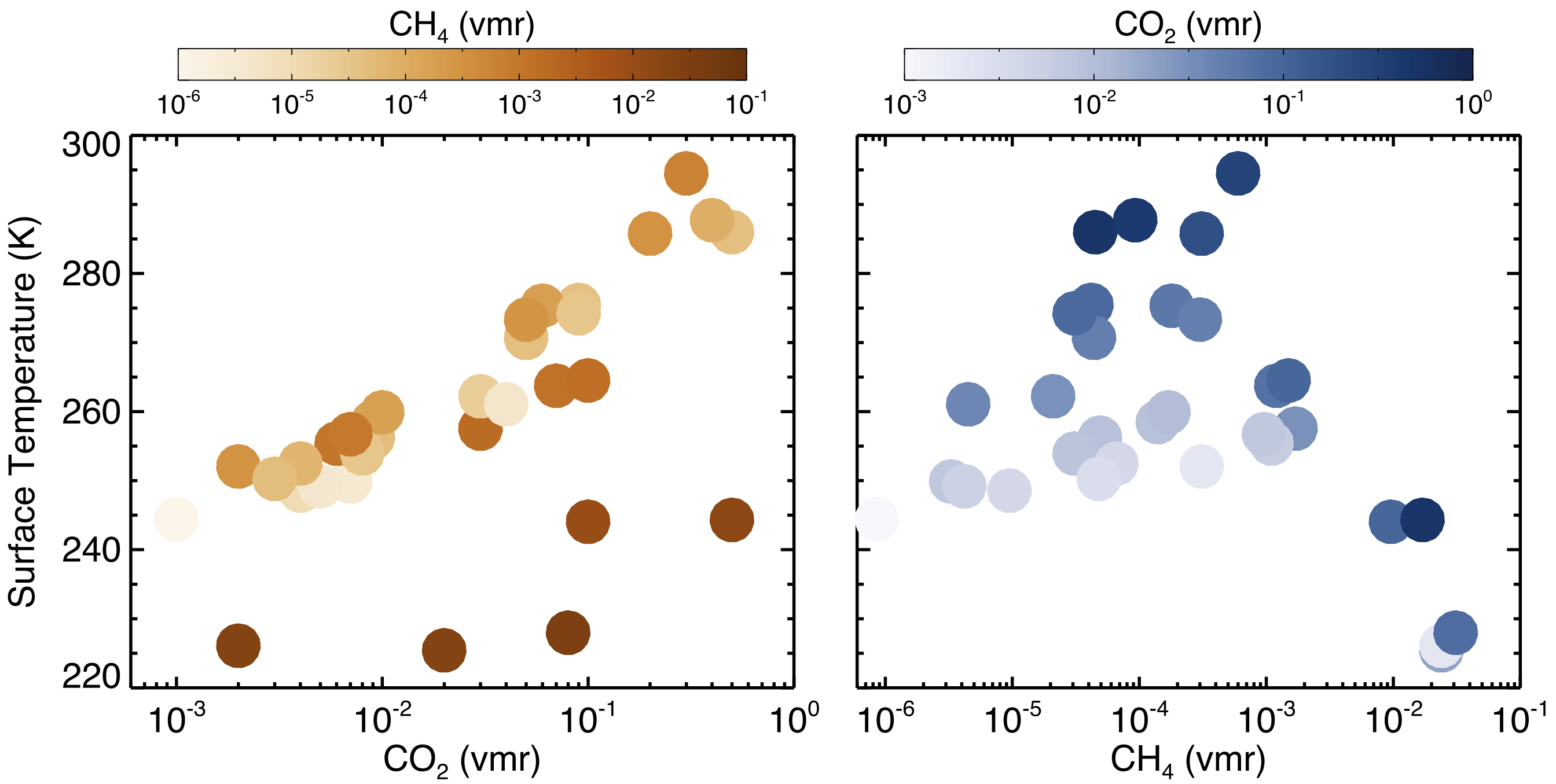

**Figure 3.** The left panel plots the global mean surface temperature vs. $CO_2$ mixing ratio for all 32 ExoCAM simulations. Circles are color-coded via their $CH_4$ mixing ratio. The right panel similarly shows the global mean temperatures, but plotted vs. $CH_4$ on the *x*-axis and with circles color-coded by the $CO_2$ mixing ratio.

Albedos and the OLR exhibit more complex variations that depend on the strength and abundance of absorbing gases, the amount and location of clouds, the temperature structure of the atmosphere, and the reflectivity of the surface. Generally, for warmer planets, the albedo decreases with increasing temperature. This is attributed to vanishing surface ice in favor of low-albedo open oceans and to increased atmospheric water vapor, which absorbs strongly in the near-infrared, particularly with incident radiation from low-temperature stars (J. Yang et al. 2016). Likewise, the OLR generally increases with increasing planetary temperature. However, for both the albedo and the OLR, we observe in Figure 4 that increasing $CH_4$ has a significant impact on the trends, with our highest $CH_4$ standing as outliers. Our highest $CH_4$ cases are our coldest cases, dominated ice coverage, while maintaining the lowest albedos and the highest values for OLR. This action is attributed to strong near-infrared absorption by $CH_4$, which is accentuated by the very red incident stellar energy from the host star, TRAPPIST-1. $CH_4$ strongly absorbs shortwave radiation in the stratosphere, limiting the energy that can reach the surface, while creating an optically thick inversion that can effectively radiate energy away to space.

To better understand the climate states revealed in our ExoCAM ensemble, now take a closer look at their atmospheres. Inspection reveals that four general representative states emerge, originating from the four corners of our greenhouse gas grid. The four states are as follows, along with the relevant case numbers from Table 2: high-$CO_2$/low-$CH_4$ (case 16), which results in a warm climate ($T_S$ = 286 K) with minimal ice coverage (11%); high-$CO_2$/high-$CH_4$ (case 18), which results in a cold ($T_S$ = 244 K) but still marginally habitable climate (88% ice coverage); low-$CO_2$/low-$CH_4$ (case 17), which also results in a cold but marginally habitable climate ($T_S$ = 249 K, 74% ice coverage); and finally low-$CO_2$/high-$CH_4$ (case 23), which results in a frozen world with photochemical hazes ($T_S$ = 227 K, 100% ice coverage). Figure 5 shows contour plots for surface temperature, total cloud water column, the top-of-atmosphere albedo, and the OLR from each of our four representative climates, respectively. Conveniently, the cases are ordered from warmest in the left column to coldest in the right column. The global mean values for each quantity are printed above the associated contour panel.

An "eyeball" climate pattern emerges (R. T. Pierrehumbert 2010), along with elevated equatorial temperatures at all longitudes due to increased atmospheric heat transport from superrotating winds (J. Haqq-Misra et al. 2018). Superrotating winds are a commonly identified feature of TRAPPIST-1 e simulations, confirmed in multiple different 3D climate models (D. E. Sergeev et al. 2020). As previously noted in reference to Figure 4, warmer planets feature a more robust hydrological cycle, leading to increased cloudiness. Warm climates in our ensemble, such as case 16 featured (also 4, 6, and 28), are dominated by open oceans and near-global cloudiness, with the thickest clouds occurring at high latitudes. For progressively colder climates, such as cases 17 and 18 featured (as well as numerous others), cloud water amounts become increasingly concentrated on the substellar hemisphere where open ocean waters remain, with a narrowing equatorial cloud belt advected eastward over the ice-covered nightside of the planet by the supperrotating winds. Lastly, frozen climates, such as case 23 featured (also 7 and 26), have very little water vapor and clouds and only immediately at their substellar point. Still, even for frozen climates a faint hydrological cycle exists, balancing light snowfall and sublimation, but only occurring where the star's energy is the strongest.

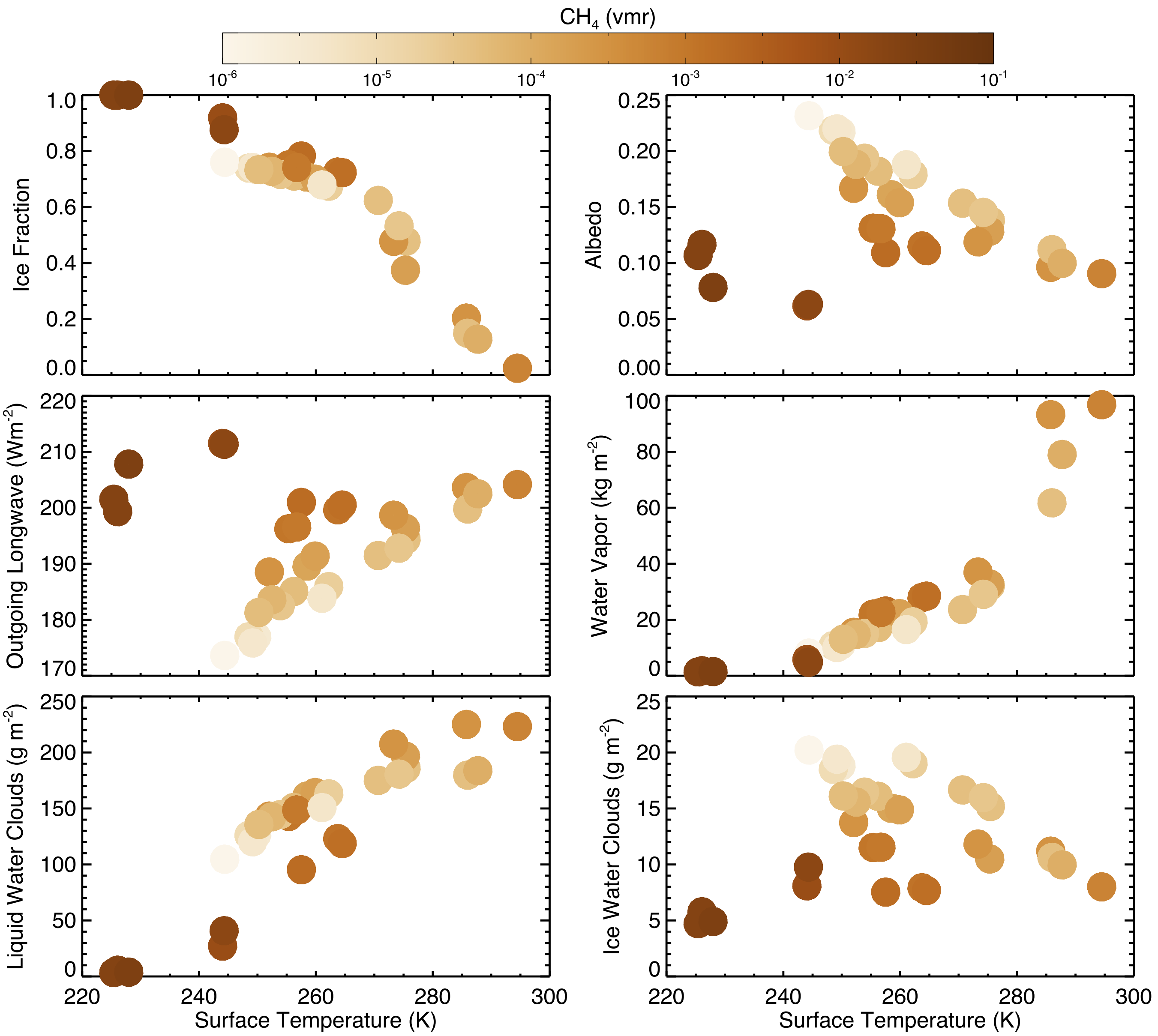

**Figure 4.** Each panel shows global mean climate quantities plotted against the global mean surface temperature on the *x*-axis, with data points color-coded by the $CH_4$ volume mixing ratio. While quantities generally display dependencies that can be linked to the surface temperature, high-$CH_4$ cases are outliers.

As noted above in reference to Figure 4, relationships governing the albedo and OLR are more difficult to discern, combining numerous interconnected factors. However, in our ensemble we find that $CH_4$ plays a significant role in determining these properties. $CH_4$ has significant absorption bands in the near-infrared, coincident with the TRAPPIST-1 spectral emission. Thus, for progressively higher $CH_4$ amounts, significant stellar radiation is absorbed by $CH_4$ in the upper atmosphere. Notably, our two high-$CH_4$ cases in Figure 4 feature relatively lower albedos and relatively higher OLR despite being the coldest simulations. Upon observing this combination of quantities on a distant world, conventional wisdom might assign these worlds as hot, cloud-free climates based on simple energy balance considerations. However, what is being observed is a stratospheric inversion driven by strong $CH_4$ near-IR absorption.

Figure 6 shows the temperature, water vapor, and cloud vertical profiles for our four representative cases. Solid lines show the global mean values, while dashed and dotted lines show the substellar and antistellar hemispheric mean profiles, respectively. As seen in Figure 5, significant day−night asymmetries in the surface temperature are found, and they are expected given the absence of ocean heat transport. However, day−night temperature gradients are considerably smaller in the upper atmospheres. Likewise, water vapor and cloud water have significant day−night hemispherical differences in the lower atmosphere, but less so in the upper atmosphere.

The vertical structure of the atmosphere reveals the interplay between the effects of $CO_2$ and $CH_4$. While both are greenhouse gases, $CO_2$ tends to cool the stratosphere, while $CH_4$ strongly absorbs in the near-infrared and thus tends to warm the stratosphere. Coupled with the red incident spectra of TRAPPIST-1, $CH_4$ can create a significant antigreenhouse effect. $CH_4$ exhibits a strong antigreenhouse effect in our simulations, causing a cooling trend for climates with $CH_4 \geqslant 10^{-3}$ (Figure 3). Similar findings have been reported in other studies performed with a variety of models, lending confidence to the robustness of this result. R. M. Ramirez & L. Kaltenegger (2018), using a 1D radiative-convective model, found that $CH_4$ can shrink the habitable zones of the coolest stars ($T_{\rm eff} = 2600$ K, comparable to TRAPPIST-1) by 10% via the $CH_4$ antigreenhouse effect. M. Turbet et al. (2018), using the LMD generic 3D climate model, found ~25 K of global mean cooling for TRAPPIST-1 e as $CH_4$ is increased from a partial pressure of $10^{-4}$ to $10^{-1}$ bars in a 1-bar $N_2$ background atmosphere. Likewise M. T. Mak et al. (2024), using the UM 3D climate model, found a similarly strong antigreenhouse effect occurring for TRAPPIST-1 e when $CH_4 \geqslant 10^{-3}$, with stratospheric inversions present even when hazes are omitted. Our results are generally in line with those of M. Turbet et al. (2018) and M. T. Mak et al. (2024). Finally, note that these prior studies used HITRAN 2012 (or earlier versions) to derive their absorption coefficients. B. Byrne & C. Goldblatt (2015) documented that the integrated strength of $CH_4$ line absorption has increased with subsequent HITRAN revisions from HITRAN 2000 to HITRAN 2012, the limit of their study. Here we used HITRAN 2016 as the basis for our correlated-*k* coefficients, which includes new $CH_4$ absorption in the near-infrared between 0.95 and 1.05 $\mu$m, a feature that is not present in HITRAN 2012 and earlier versions (see I. E. Gordon et al. 2017, Figure 14). The new $CH_4$ absorption band is coincident with the peak in emission with the TRAPPIST-1 spectra, lending to an even stronger impact of the $CH_4$ antigreenhouse effect found in this work.

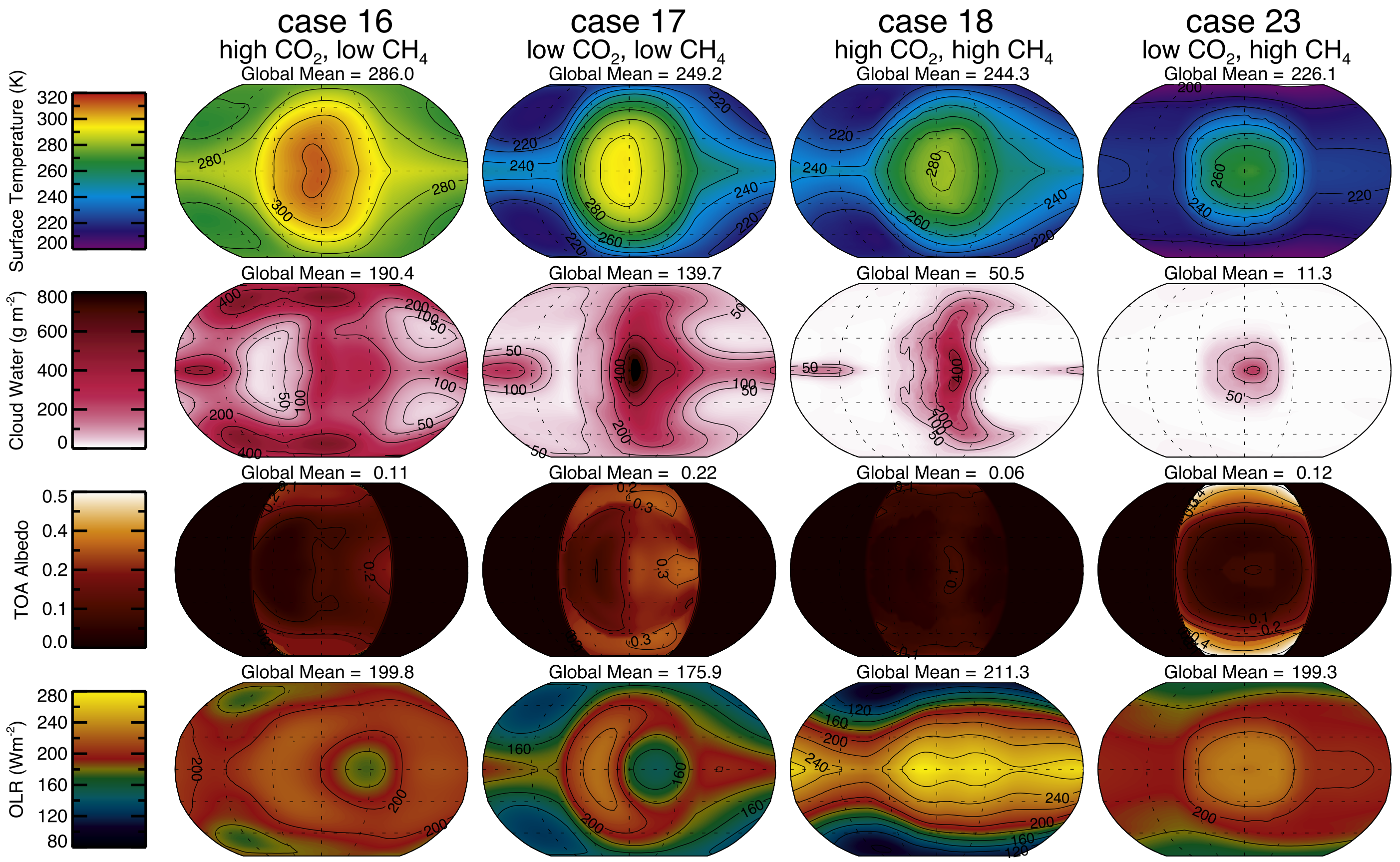

**Figure 5.** Contour plots of surface temperature, cloud water column, albedo, and OLR for our four representative climate states.

For the three cases (7, 23, 26), `Atmos` predicts that optically significant photochemical hazes will form. For these cases we use `CARMA` coupled to `ExoCAM` to model fractal aggregate hazes. Haze production rates as a function of altitude and zenith angle are taken directly from `Atmos` and are input into `ExoCAM-CARMA` as a mass source function for haze particles. Fractal aggregate microphysical assumptions follow directly from E. T. Wolf & O. B. Toon (2010). Hazes are only found in atmospheres with the highest $CH_4$, which are correspondingly the coldest atmospheres owing to strong $CH_4$ antigreenhouse cooling. In our three hazy cases the global mean haze production rates calculated by `Atmos` are $3.08 \times 10^{14}$ g, $4.35 \times 10^{14}$ g, and $2.02 \times 10^{14}$ g per Earth year for cases 7, 23, and 26 respectively. Note that the haze production is given in terms of an Earth year as an artifact of `CARMA`'s native linkage to `CESM` for Earth studies. Our values for haze production are in line with laboratory estimates of haze production rates for the Archean Earth (M. G. Trainer et al. 2004, 2006). When in equilibrium, the total haze mass present in the atmosphere is $5.77 \times 10^{14}$ g, $6.07 \times 10^{14}$ g, and $3.915 \times 10^{14}$ g for cases 7, 23, and 26, respectively. It follows that the residence time of haze aerosols is 1.87, 1.40, and 1.94 Earth years for each of the three hazy cases. Residence times vary inversely with the production rate, with higher production rates leading to larger particles that fall out of the atmosphere faster.

In our study, we find that hazes have very little effect on the surface climates. For these low-$CO_2$/high-$CH_4$ cases, the surfaces are already very cold owing to a limited $CO_2$ greenhouse and strong $CH_4$ antigreenhouse cooling. When duplicated with hazes omitted, the global mean surface temperatures for cases 7, 23, and 26 are 229.1, 228.4, and 231.4 K, respectively, representing only a couple of degrees of cooling caused by the haze in each case. In Figure 7 we plot vertical profiles of temperature, haze mass, and haze particle effective radii. In the temperature panel, solid lines show the hazy cases, while the dashed lines show the results from duplicate simulations with the haze artificially omitted from the simulations. In the radius panel, the solid lines illustrate the radii of fractal particles, while the dashed lines refer to the equivalent spherical radius, which is the radius the particles would have if artificially condensed into uniform spheres.

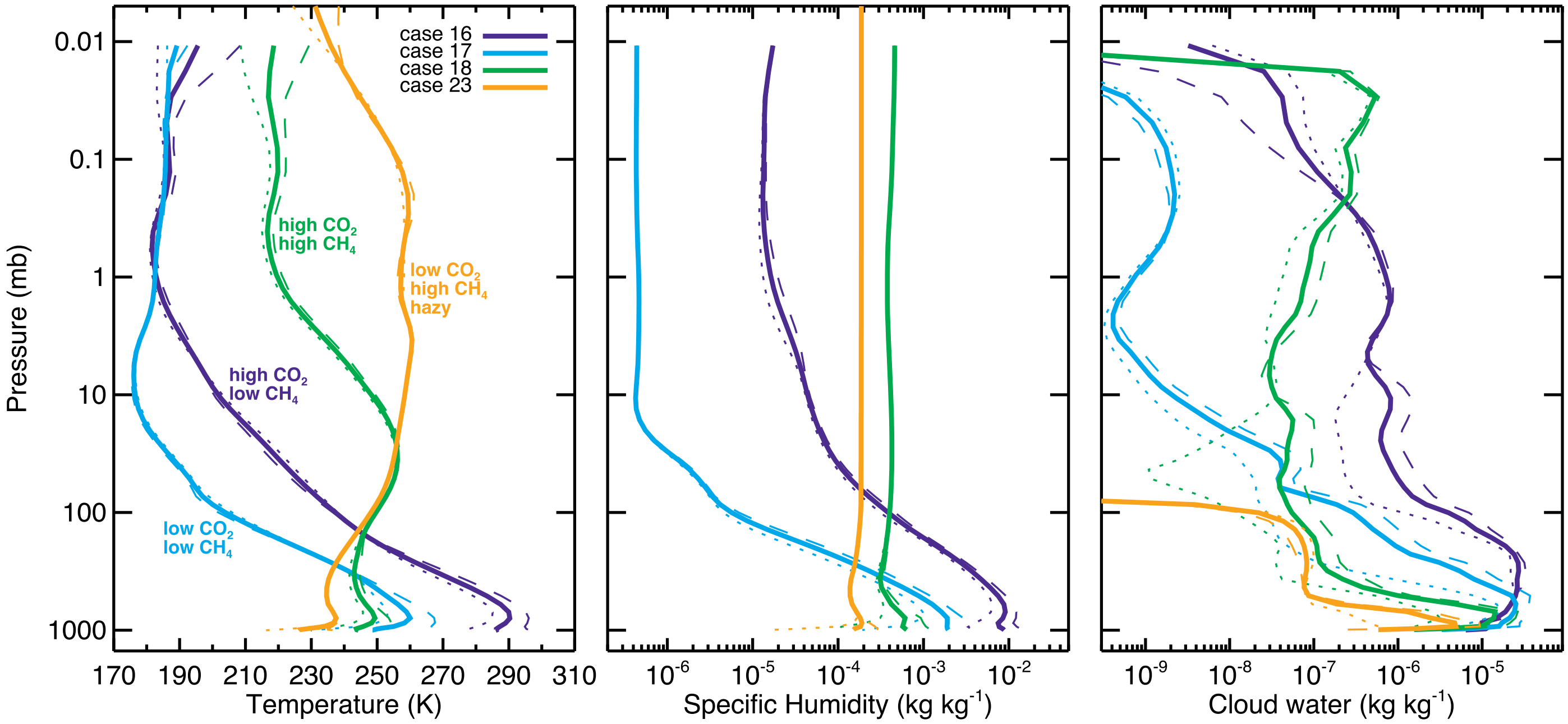

**Figure 6.** Vertical structure of temperature, specific humidity, and cloud water for our four representative cases. Note that case 23, which includes haze, was run with a higher model top than the nonhazy cases. Solid, dashed, and dotted lines show the global mean, substellar hemisphere mean, and antistellar hemisphere mean, respectively.

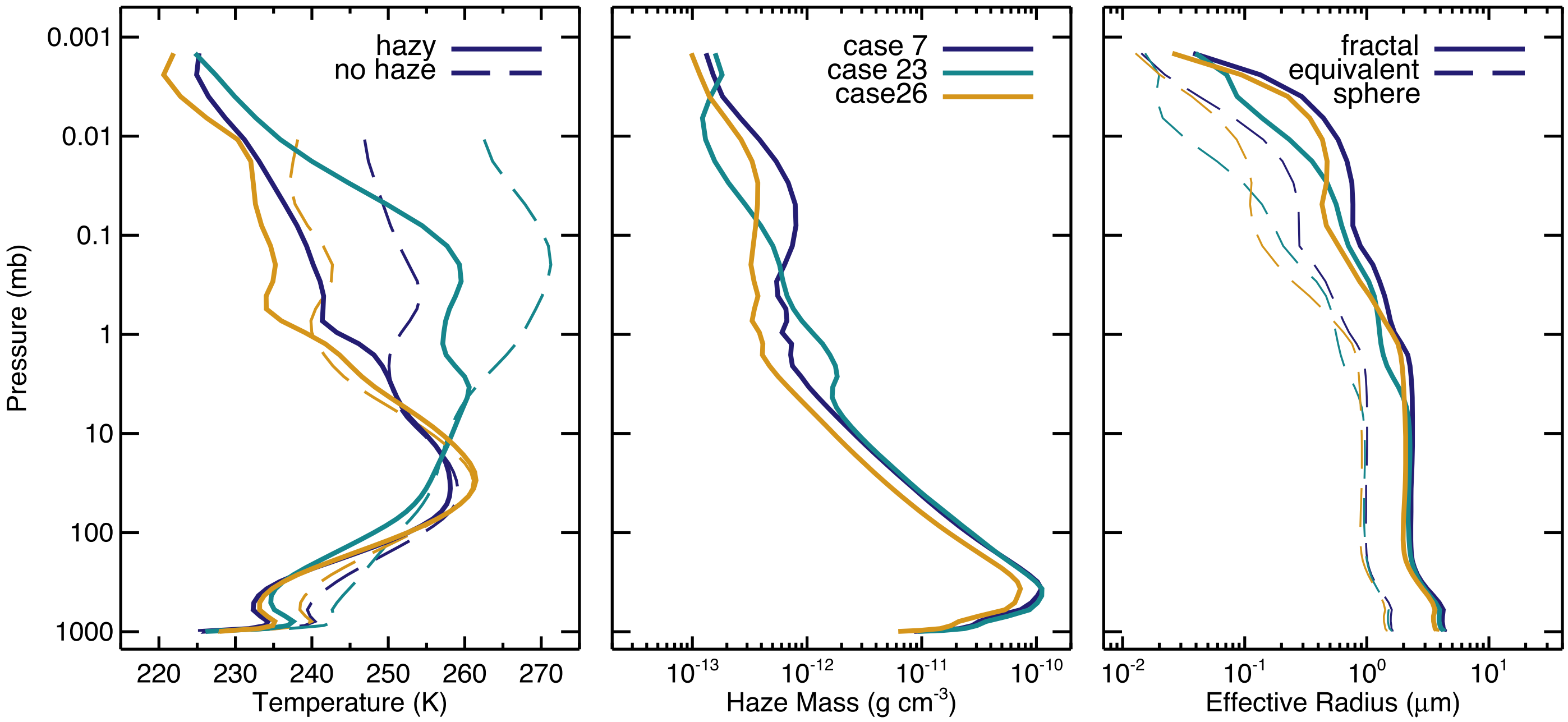

**Figure 7.** Vertical profiles of temperature, haze mass, and haze effective radii for our three hazy cases (7, 23, 26). In the left panel, dashed lines indicate the temperature profile when no haze is assumed to form and with all else being equal. In the right panel, solid lines show the fractal aggregate radii and dashed lines show the radii if each aggregate were condensed into a solid sphere.

A surprising feature that stands out in Figure 7 is that the simulations with haze have stratospheres that are colder by 10–15 K compared to their haze-free counterparts. We suspect that this is due to the particular radiative assumptions made in our haze simulations, along with consequences of the 3D distribution of hazes in the atmosphere. Figure 8 shows the TRAPPIST-1 spectra compared with the haze absorption coefficient for a $\sim$1 $\mu$m fractal aggregate and an equivalent-mass Mie particle. Note from Figure 7 that fractal aggregate particle sizes are about 1 $\mu$m coincident with the stratospheric temperature anomaly. We see that the haze absorption coefficient drops sharply longward of $\sim$0.6 $\mu$m, with a broad minimum in absorption existing from 1 to 3 $\mu$m, precisely lining up with where TRAPPIST-1 emits most of its radiation.

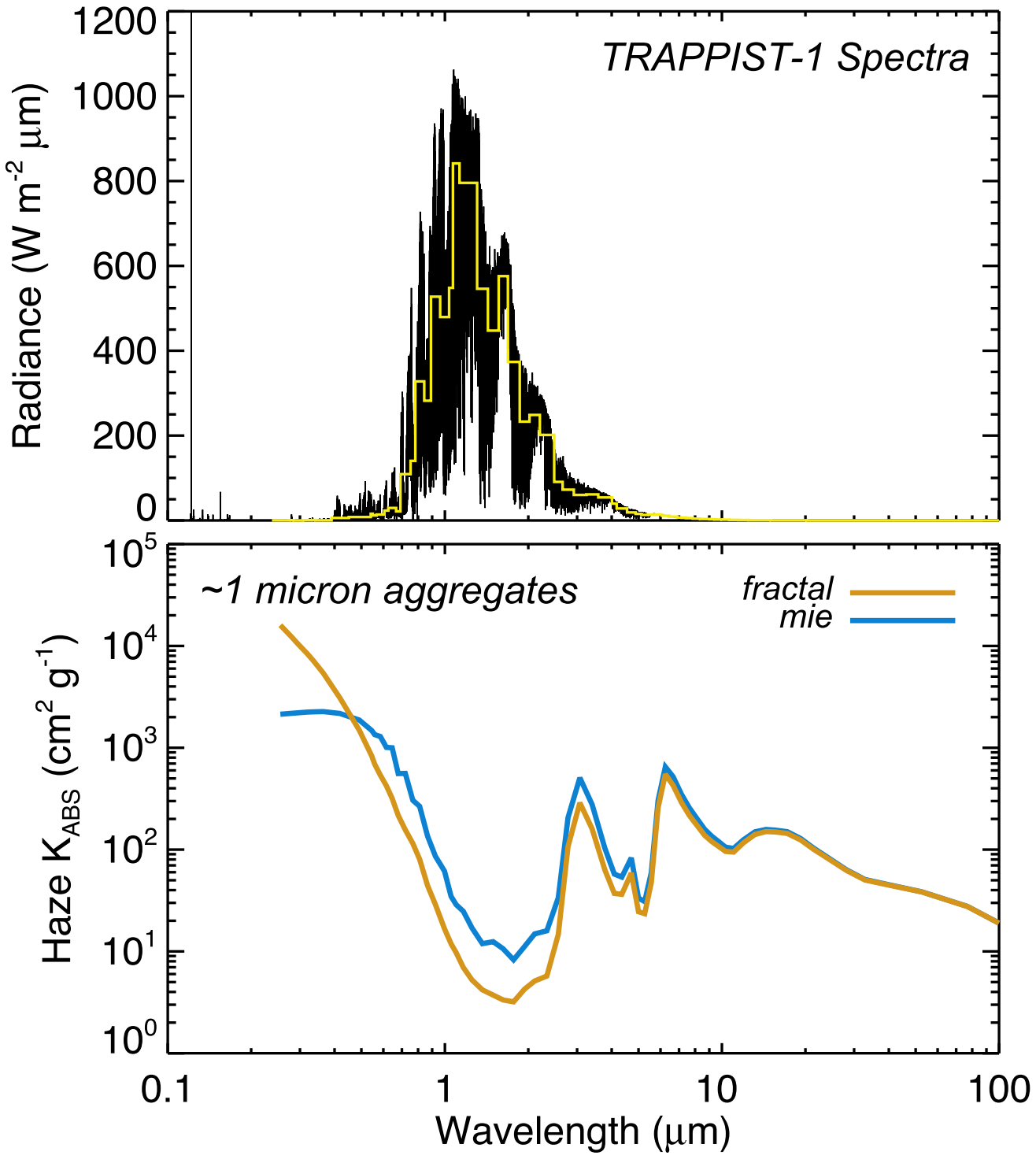

**Figure 8.** The TRAPPIST-1 spectral energy distribution (top panel) compared against the haze mass absorption coefficient for a ∼1 μm particle (bottom panel). The light-yellow histogram in the top panel shows the GCM gridded stellar spectrum. The bottom panel illustrates differences between the optical properties of fractal aggregates vs. equivalent-mass Mie particles.

The fractal nature of the particles deepens the minimum slightly compared to equal-mass Mie particles. The broad minimum itself is a direct reflection of the spectral structure of the imaginary refractive indices reported by B. N. Khare et al. (1984, see their Figure 4), which are used here.

Furthermore, the general circulation dictates that the bulk of the haze mass accumulates in areas of the planet with low stellar irradiance. Figure 9 shows the total atmosphere haze column mass and the stratospheric haze column mass for case 23. For the calculation in Figure 9, we consider the stratosphere to be 1 mbar and above. Combining the information of Figures 7 and 9, it is evident that most of the haze mass resides in the polar troposphere at latitudes ⩾60°. A similar result was found in E. T. Wolf & O. B. Toon (2010) for the early Earth. In the stratosphere, the bulk of the haze mass accumulates on the nightside of the planet, at and beyond the eastern terminator. Past 3D studies similarly found that dayside-to-nightside stratospheric circulations may dictate $O_3$ distributions, indicating that this may be a general feature of tracer transport on tidally locked planets (M. Braam et al. 2023). The combination of having little haze subject to direct stellar irradiance on the day hemisphere and the spectral coincidence of TRAPPIST-1's maximum in emission and B. N. Khare et al. (1984) haze refractive indices' minimum in absorption mutes any effect of shortwave heating driven by the presence of haze.

At longer wavelengths, i.e., those typical of thermal emission from temperate planets, haze absorption coefficients increase (Figure 8). The enhanced thermal absorption properties lead in kind to increased emissivities and thus enhanced radiative cooling rates to space, much like how $CO_2$ cools planetary stratospheres. Indeed, we find a maximum in radiative cooling rates coincident with the nightside stratospheric haze mass concentration illustrated in Figure 9. In summary we find that $CH_4$ near-infrared absorption of stellar energy drives the observed stratospheric inversions while hazes do not contribute appreciably to heating owing to their spectral properties and location, but hazes do contribute to enhanced stratospheric cooling from the night hemisphere. We caution that the results are sensitive to the stellar energy distribution and refractive index combination.

### *3.4. Theoretical Transmission Spectra*

Exploring theoretical exoplanet chemistry and climate is academically interesting and perhaps even fun. However, verifying the presence of an expected atmospheric composition and climate requires connecting them to observations. In this new era of the James Webb Space Telescope (JWST), meaningful observations of potential terrestrial exoplanet atmospheres are beginning to be reported (T. P. Greene et al. 2023; N. Madhusudhan et al. 2023; S. Zieba et al. 2023; C. Cadieux et al. 2024; M. Damiano et al. 2024), and observations of TRAPPIST-1 and its planetary system are in progress. We are soon approaching a time where our modeling efforts will be useful both for direct comparison against observations from specific targets and for making more general predictions across common parameter spaces. In this section, we calculate theoretical transmission spectra from our ensemble of TRAPPIST-1 e simulations using the `GlobES` module of the `Planetary Spectrum Generator`. Time-averaged atmospheric profiles from the terminator region produced by `ExoCAM` are input to `GlobES` with all relevant gases, clouds, and hazes. Figure 10 shows transmission spectra calculated from our four representative cases described in Section 3.3. The atmospheric properties of these four cases are listed in Table 2 and illustrated in Figures 5 and 6.

The four cases shown in Figure 10 represent pseudo-endmember instances of our `ExoCAM` ensemble. Spectral regions for $CO_2$ and $CH_4$ are marked at the bottom of the plot. The relative strength of each gas absorption line generally scales with its abundance, with $CH_4$ displaying greater sensitivity to abundance changes than the 4.3 μm $CO_2$ line. The continuum level for each case is set by the abundances of clouds and haze. Warmer cases, like case 16, have significantly more water clouds than do the cooler cases in the ensemble, and thus its transmission spectra have an elevated continuum level compared to the cooler cases 17 and 18. Cold cases with high $CH_4$ (case 18) display sharp $CH_4$ features juxtaposed against a low continuum level. However, the hazy case (case 23) stands out among them all, with a significantly raised and sloped continuum level.

Figure 11 highlights the effects of haze on the transmission spectra of case 23. Shown are the standard case with fractal haze, along with a Mie haze assumption, and finally one version with hazes ignored completely. If hazes are artificially turned off, the amplitude of $CH_4$ features can approach ∼100 ppm, due to exceedingly high methane abundances. However, with haze self-consistently included in the simulations, many $CH_4$ features get washed out, particularly those shortward of 2 μm. While the absolute transit depth of the 4.3 μm $CO_2$ line is largely unaffected for each instance, hazes significantly reduce its relative transit depth by increasing the continuum to the altitude of the optically thick haze deck. While it is discouraging that hazes (and water clouds to a lesser extent) can mute or outright hide underlying transmission spectral features, therein lies an opportunity to characterize exoplanetary atmospheres through differentiating the slopes of observed shortwave continuum. Note in Figure 11 that a Mie haze has a different slope than does a fractal haze. As highlighted by M. G. Lodge et al. (2024), different haze assumptions for refractive indices, fractal properties, and the optical model will all have significant effects on the nature of the haze continuum. This can also be seen in Figure 12, where the Henyey–Greenstein phase functions at 0.86 $\mu$m for fractal and Mie hazes are compared for a particle effective radius of 0.5 $\mu$m. Fractal particles have a more complex and irregular structure compared to the smooth, spherical particles assumed in Mie theory. The irregularity of fractal particles reduces the efficiency of forward scattering and enhances the scattering at other angles. Therefore, more photons are removed from the direction of propagation of the transmitted light, leading to a greater opacity. Differentiating between the Rayleigh scattering slope of different gases, flat slopes of water clouds, and the sharper slope of hazes or other aerosols could provide an interesting avenue for characterizing exoplanetary atmospheres.

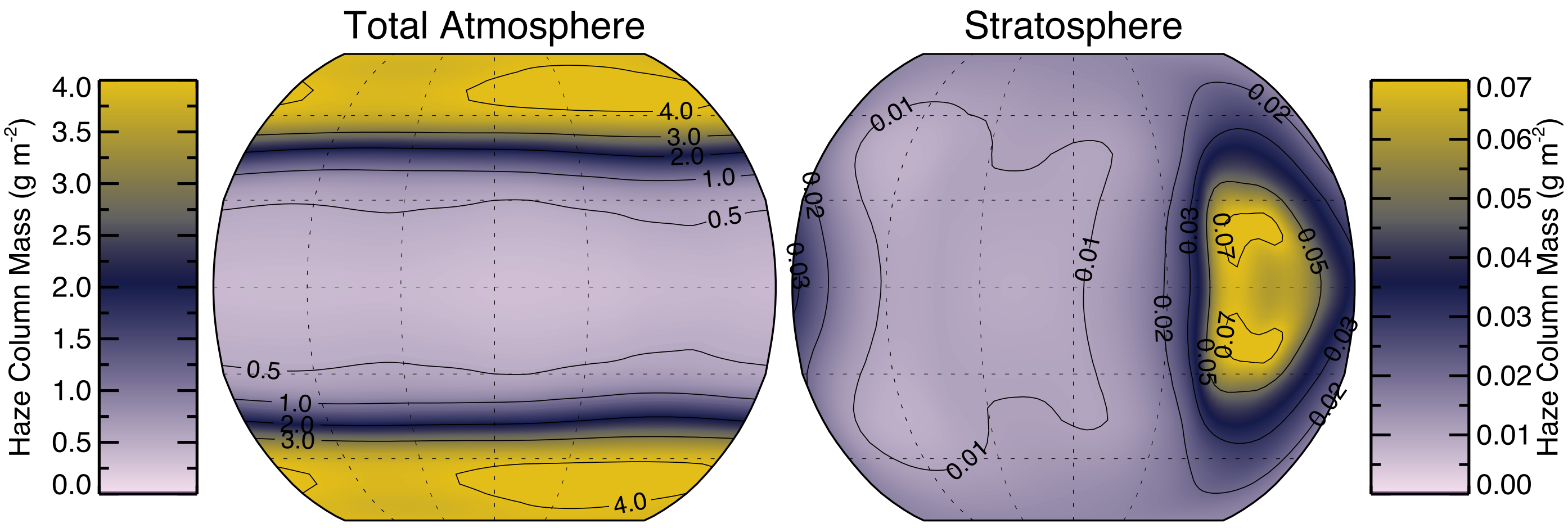

**Figure 9.** Haze column mass from case 23. The left panel shows the haze mass integrated through the whole atmospheric column. The right panel show the haze mass integrated from 1 mbar to the top of the model.

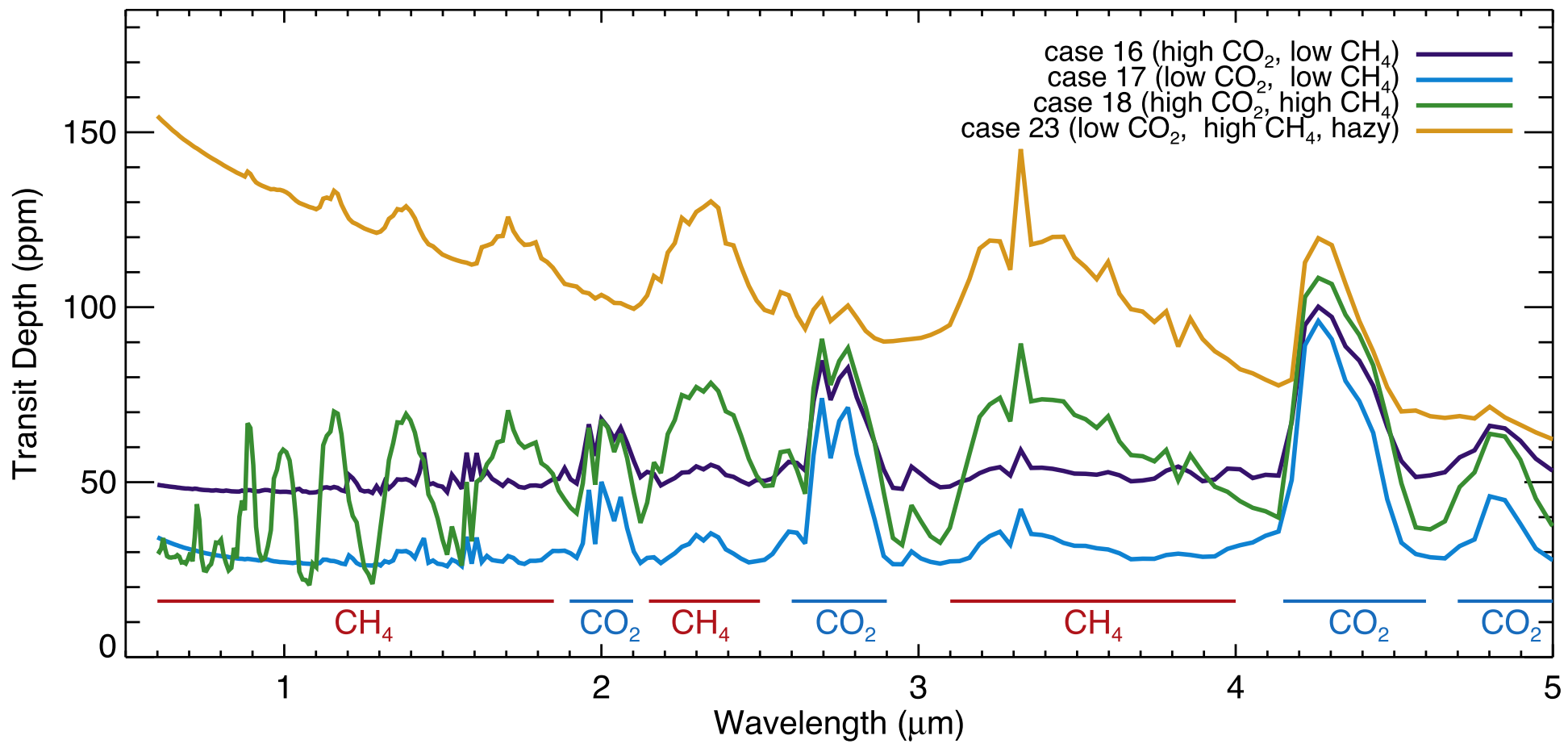

**Figure 10.** Transmission spectra for our four representative cases.

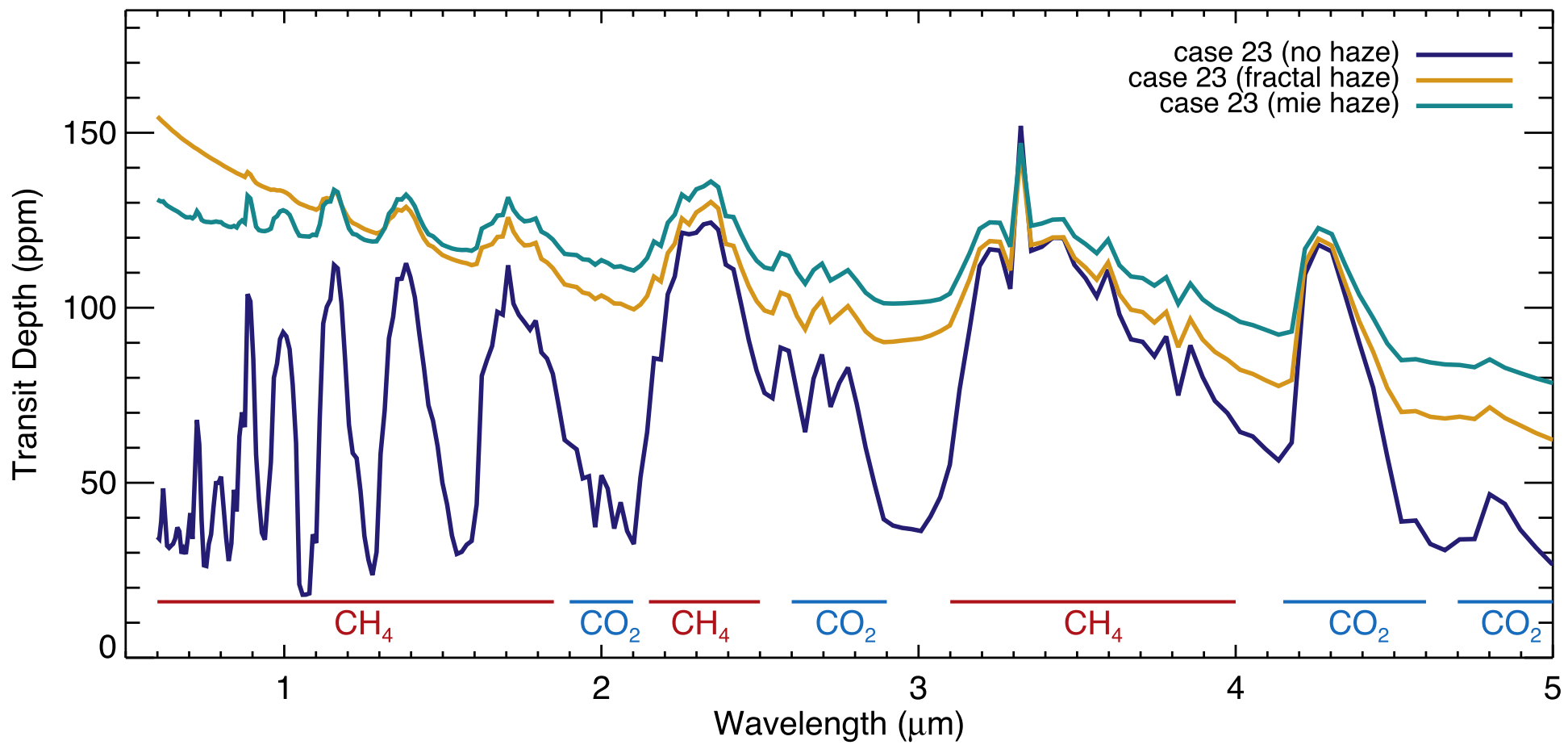

**Figure 11.** Transmission spectra for case 23 considering fractal aggregate haze, Mie haze, and no haze.

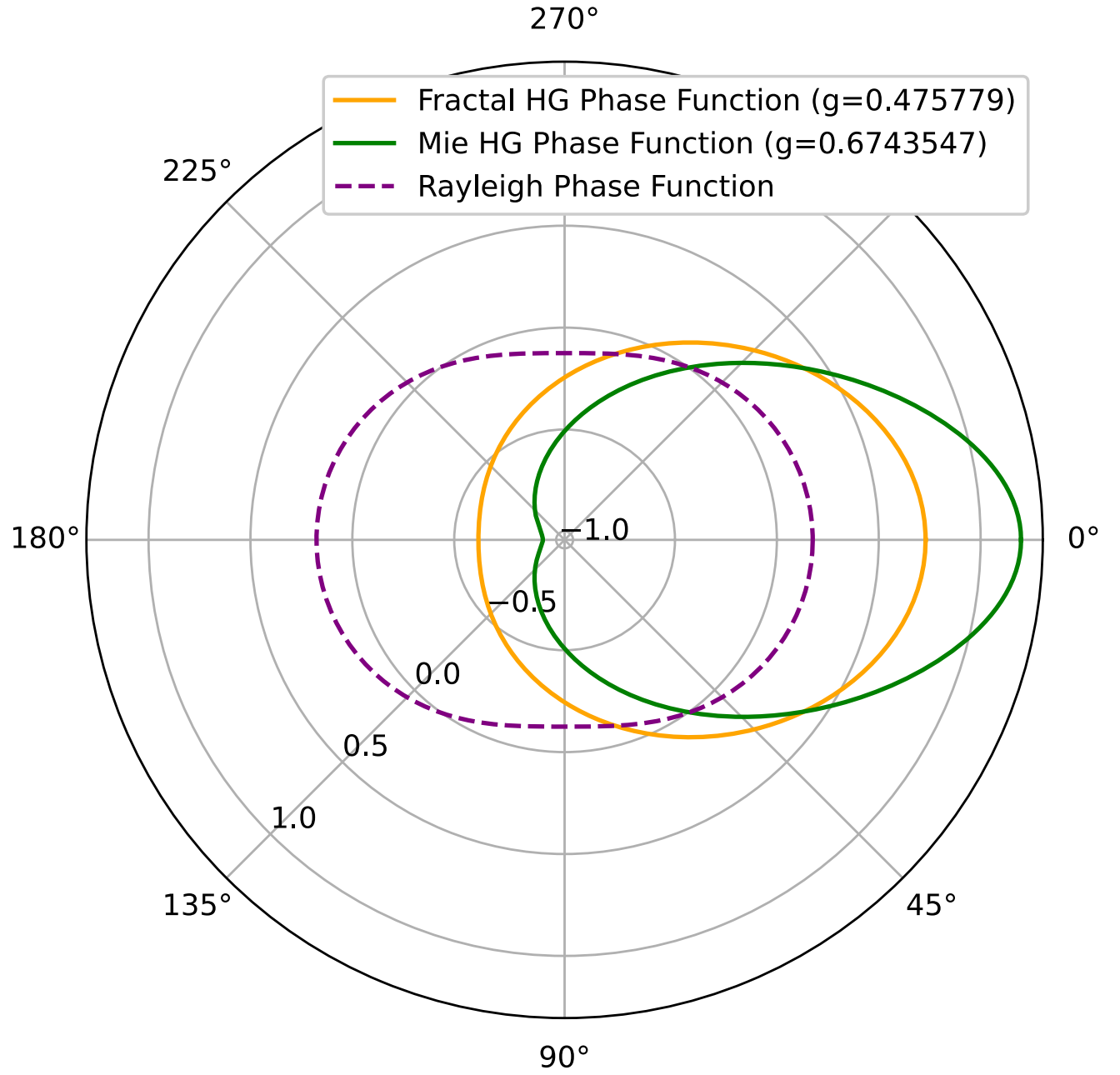

**Figure 12.** Polar plot showing the Henyey–Greenstein scattering phase function at 0.86 $\mu$m for the fractal (orange) and Mie (green) haze optical properties assuming an effective radius of 0.5 $\mu$m. The Rayleigh scattering phase function (purple dashes) is also display for comparison.

## 4. Synthesizing the Grid

Computing a grid of simulations (sparse or otherwise) and reporting on what our models regurgitate is a straightforward endeavor. In Sections 3.1, 3.3, and 3.4 we have done just that, reporting on the basic emergent features of chemistry, climate, and transmission spectra of TRAPPIST-1 e with varying $CO_2$ and $CH_4$ compositions. However, recall that one of our primary goals in this study is to demonstrate the utility of sparse sampling approaches toward exoplanet climate problems. Synthesizing the sparse sampled model results into a coherent whole proved to be the most challenging step in our process, requiring nuance and cleverness. In our early attempts to synthesize the sparse sampled model results, we naively relied on traditional interpolation approaches. However, the sparseness of our 32-model `ExoCAM` sample, lack of any regular grid underlying the QMC distribution of the sample, and the inherent nonlinear feedbacks of climate systems resulted in renderings across our 2D parameter space that were noisy to the point of being indecipherable. After trial and error, we settled on using kriging to interpolate across our sparse grid, following the description in Section 2.3. The ordinary kriging algorithm depends on the choice of a "variogram" function that gives a statistical relationship between the spatial separation and dissimilarities of data points across the domain of interest (which is comparable to the "kernel" function used in Gaussian process emulation (see, e.g., H. Wackernagel 2003; J. Haqq-Misra et al. 2024; M. Marinescu 2024). We use a linear variogram function for all the kriging interpolations of global mean quantities shown in this section, with the exception of surface temperature (using an exponential variogram function) and column-integrated water vapor (using a spherical variogram function). Using this technique gives us an exact interpolation at all our simulated points that would not be obtainable with other conventional interpolation methods. Our use of ordinary kriging provides a reliable interpolation across our sparse grid. This approach produces useful information content through which to interpret our ensemble results and will guide any future calculations toward the most interesting regions of the parameter space.

In Figures 13, 14, and 15 we have used kriging to interpolate climate and observable properties from our 32 `ExoCAM` and `PSG` simulations, across the full $CO_2$ and $CH_4$ parameter space first described in Section 3.1. Note that in these parameter space synthesis plots we use $CO_2$ mixing ratio as the $y$-axis and $CH_4$ mixing ratio as the $x$-axis, as opposed to $CH_4$ surface flux as the $x$-axis as seen in Figures 1 and 2. We feel that plotting against $CH_4$ mixing ratio is more intuitive than plotting against $CH_4$ surface flux, as it directly relates to the amount of gases present in the atmosphere. In all plots, note that the kriging technique produces smooth contour plots from our sparse grid points, which we consider as a reliable representation of the variables of interest across the parameter space. We do note that some artifacts appear in places were few data points exist, which can suggest areas of the parameter space for further exploration. Nevertheless, such artifacts remain limited in scope and do not appear to hinder our analysis of the broad trends across this parameter space.

Because our application of ordinary kriging is designed to yield exact interpolation at our simulated quantities, we do not have easily quantifiable errors to assess the accuracy of our interpolations. (This is in contrast to use cases of Gaussian process emulation that are intended for "emulating" a model across a parameter space to minimize total error, rather than to obtain exact interpolation at the data points; see, e.g., R. B. Christianson et al. 2023.) For this reason, we have conducted our sampling using a series of three Sobol sequences (shown in Figure 2), which allows us to calculate the percent error from one of these sequences when compared with the kriging interpolation obtained from another. As a visual example, the different kriging interpolation patterns for global mean surface temperature are shown in Figure A1, when considering sequence 1, sequence 2, and sequence 3 in isolation, as well as the combined sequence 1+2. Visual inspection shows that the general patterns remain consistent across the four panels, although notable differences are also apparent; these figures can all be compared with the combined plot of global mean surface temperature using all three sequences (Figure 13, top left panel). We have also quantified the percent error for all our global mean quantities in Table A2, which includes comparisons of sample 2 data points to the sample 1 interpolation, sample 3 data points to the sample 2 interpolation, and sample 3 data points to the sample 1+2 interpolation. In general, the errors all tend to reduce as more data points are used for kriging interpolation, as expected. The largest errors occur for the cases with haze, as well as those with high $CH_4$ abundances, as these have the greatest nonlinear behaviors compared to the low-$CH_4$ and haze-free cases. Percent errors for surface temperature tend to be relatively low for most cases (around 1% to a few percent), while errors for cloud quantities tend to be slightly higher (from a few percent to 10% for most cases) owing to the nonlinearity of cloud behavior. For estimating the number of transits required for 5$\sigma$ detection, the range of percent errors is larger (a difference of a few to several tens of percent in most cases), with larger variation for $CH_4$ than $CO_2$; this behavior is also consistent with the strong nonlinear behavior that is expected when combining climate model output with spectral detection models. This quantification of errors is intended to illustrate that our application of kriging is generally robust and to highlight the regions of parameter space likely to carry the greatest degree of uncertainty. However, it is also important to note that the percent errors listed in Table A2 do not represent the actual errors from our global mean quantities output by our models and shown in Figures 13, 14, and 15, because all of these figures show exact interpolated quantities. Calculating errors for these interpolations would require us to simulate yet another set of GCM runs (using another Sobol sequence) and then compare these with our interpolations. Such a process may be useful if the ultimate goal is to create an emulator; however, our goal in this study is to understand the behavior of climate across our parameter space of interest using a sparse sample. We have also included contour plots showing the standard deviation (square root of the kriging variance) across the parameter space for all the kriged quantities, which are shown in Figures A2, A3, and A4. Unsurprisingly, the standard deviation around the sampled points is low, and the standard deviation increases as distance from a sampled point increases. It is important to note that the kriging standard deviations shown are not absolute estimates of error, but rather give an appreciation of the varying precision of kriged quantities (H. Wackernagel 2003). To calculate true error rates, we would need to conduct an additional set of GCM simulations in the gaps in our parameter space to consider as truth and evaluate against the emulated values. We recognize that further extension of our sample could decrease the uncertainty of the interpolated space, and we will leave it for future work to explore specific regions of this parameter space at finer detail.

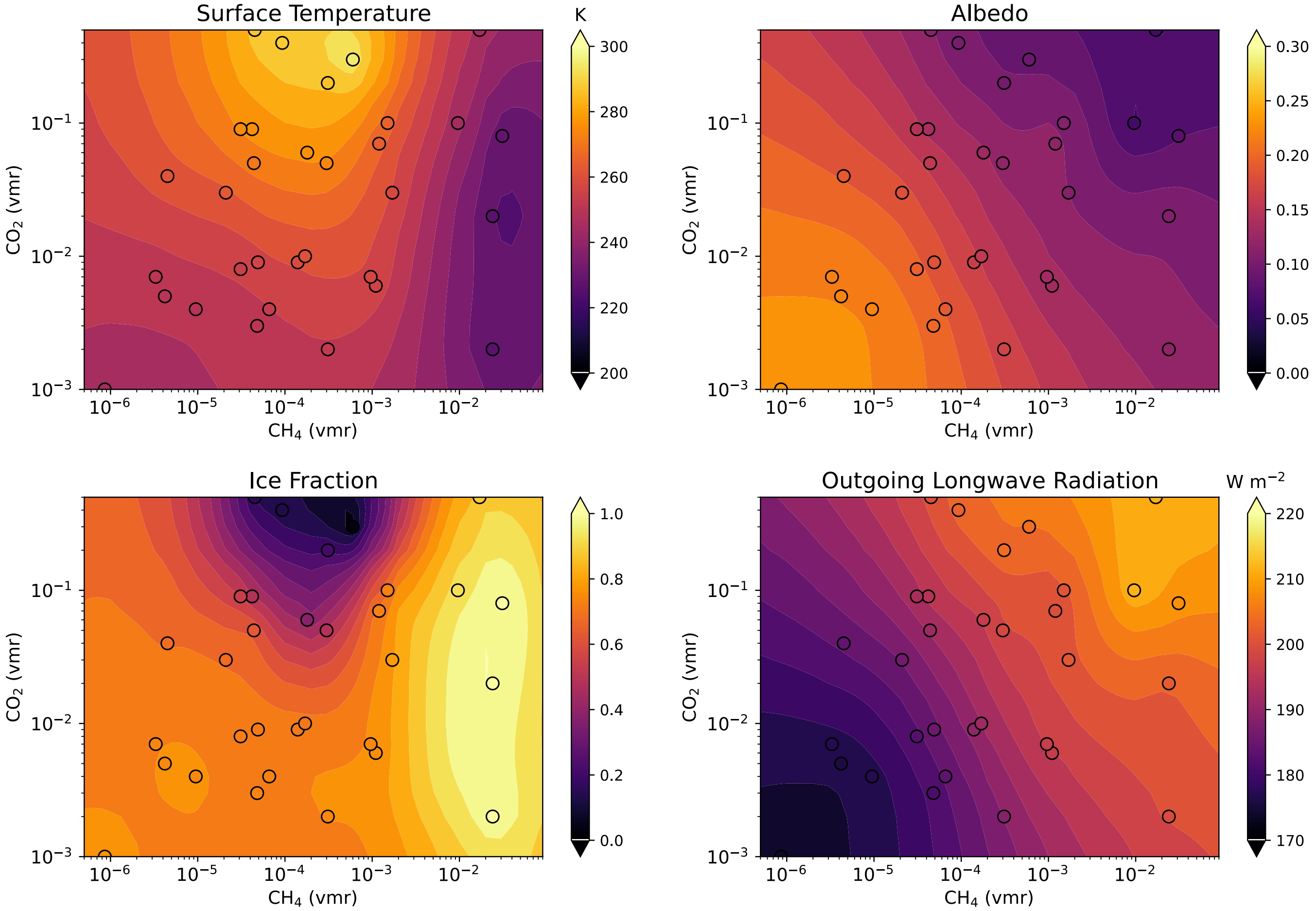

**Figure 13.** Kriging interpolation of the 32 ExoCAM simulations. The black circles show the ExoCAM simulations. Panels are shown and labeled for the global mean surface temperature, albedo, ice fraction, and OLR.

With kriging interpolation completed, trends and relationships among different climate quantities are more readily understood by simple visual inspection. Figure 13 shows the results for surface temperature, albedo, sea-ice fraction, and OLR. As discussed some in Section 3.3, there is a balance between $CO_2$ and $CH_4$. $CO_2$ is a pure greenhouse gas, with limited effect on stellar radiation in the regimes studied here ($CO_2 \leqslant 0.5$ bars). As such, increasing $CO_2$ warms the planet in virtually all instances. Note that the sea-ice fraction naturally mirrors the global mean surface temperature. $CH_4$ is a greenhouse gas but also features significant near-infrared absorption that coincides with the peak in TRAPPIST-1 spectral emission. High amounts of $CH_4$ also tend to yield photochemical haze formation. In Figure 13, we see that there is a limit to climate warming possible for TRAPPIST-1 e via $CH_4$. Warming is observed for increasing $CH_4$ mixing ratio up to $\sim 10^{-3}$. However, for higher values, the climate cools owing to the strong near-IR absorption and the resultant antigreenhouse, significant for planets orbiting red stars (A. D. Del Genio et al. 2019; T. J. Fauchez et al. 2019). Haze formation does not have a strong effect on the climate results, as was previously discussed.

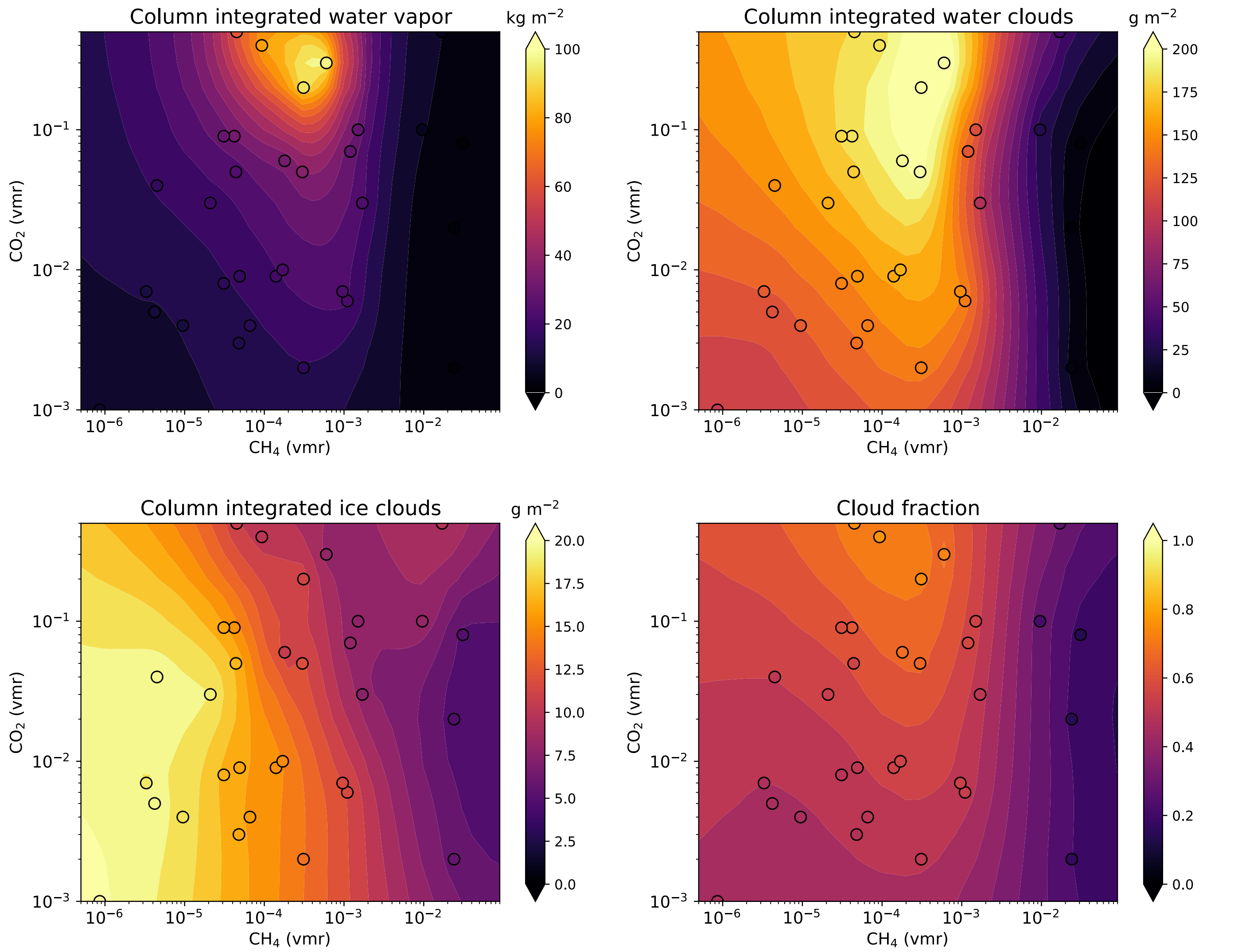

**Figure 14.** Kriging interpolation of the 32 ExoCAM simulations. The black circles show the ExoCAM simulations. Panels are shown and labeled for the column-integrated water vapor, liquid water clouds, ice water clouds, and cloud fraction.

The albedo and OLR display more complicated correlations. Planets in the lower left quadrants have low greenhouse gas amounts and generally are cold with high ice fractions and have high albedos and low OLR, a presentation that would agree with commonly held intuitions about planetary climate. In the upper right quadrant planets are even colder and icier but feature an opposite presentation, having low albedos and high OLR. This apparent oddity is due to strong $CH_4$ absorption in the stratosphere, which simultaneously lowers the albedo, cools the surface by preventing light transmission, and heats the stratosphere, creating an optically thick inversion layer that is able to efficiently radiate to space, raising the emitting temperature of the planet beyond that of the surface itself. Our results hint at how disk-averaged emission and albedo quantities from a given exoplanet could easily fool us.

Figure 14 shows the results for the column-integrated water vapor, liquid water clouds, ice water clouds, and cloud fraction. Water vapor and clouds of course play significant roles in the climate system, with the water vapor greenhouse effect amplifying temperature perturbations and clouds having competing effects, contributing to both the greenhouse and the albedo. However, clouds also significantly affect transmission spectra via modulating the continuum level. In Figure 14 we see that the water vapor abundance naturally tracks with the overall planetary temperature in accordance with the Clausius–Clapeyron relation. Liquid water clouds and cloud fraction also mostly mirror the surface temperature mapping shown in Figure 13, with warmer climates being wetter and cloudier. Ice clouds are more prevalent for cool climates found at the lower left of the panels. For snowball climates at the right of the panel the water vapor and clouds fall to minimal values.

Lastly, our fundamental goal is to connect atmospheric compositions to climate states to observations made with JWST. Figure 15 shows heat maps for detecting $CO_2$ and $CH_4$ across our grid. Plotted is the number of transits required for a $5\sigma$ detection of $CO_2$ and $CH_4$. Note that we do not show an identical plot for $H_2O$ detections because the number of transits required is 1–2 orders of magnitude larger than for $CO_2$ and $CH_4$ and would be infeasible with JWST, due to reasonable time allocation constraints. For these calculations we used PSG with the methodology established by T. J. Fauchez et al. (2022) for simulating JWST transit observations with the Near Infrared Spectrograph (NIRSpec) Prism mode (S. Birkmann et al. 2022), including noise and instrumental effects. We assume a clear filter with the SUB512S subarray and employ the rapid readout pattern, specifying two groups per integration with a frame time of 0.225 s. This observing mode leads to partial saturation near the peak of the stellar energy distribution, aligning with the results reported by N. E. Batalha et al. (2018) and J. Lustig-Yaeger et al. (2019). A transit duration for TRAPPIST-1 e of 3345 s (0.93 hr) was adopted based on E. Agol et al. (2021). To incorporate noise contributions from the out-of-transit baseline, we referenced the JWST NIRSpec GTO proposal 1331 (N. T. Lewis et al. 2018), which designates around 4 hr for each transit observation. Consequently, the total out-of-transit observation time is approximately three times the transit duration. We then stack our simulated noisy transit observations until we reach our desired $5\sigma$ detection thresholds for $CO_2$ and $CH_4$, considering the signal from each gas integrated across the entire NIRSpec range (0.6–5.3 $\mu$m). Here $\sigma$ refers to the confidence interval for which gas absorption signals can be differentiated from random fluctuations in the measurement system. A $5\sigma$ confidence interval implies that the gas absorption signal deviates from random noise by 5 standard deviations. In other words, there is a 99.99994% probability that the signal is indeed genuine. At a given wavelength, a $5\sigma$ detection threshold implies that the signal is 5 times larger than the noise. For instance, here the 4.3 $\mu$m $CO_2$ feature has a typical depth of ~60 ppm (Figure 10), implying that the noise is ~12 ppm to satisfy the $5\sigma$ requirement. The noise is a wavelength-dependent quantity, and we have integrated across the entire spectrum to maximize the information content. Note that the use of a $5\sigma$ threshold for affirming observations is advocated for in the search for biosignature gases on extrasolar worlds (V. Meadows et al. 2022). Note that we do not consider observations by the JWST Mid Infrared Instrument, due to its higher noise characteristics, resulting in significantly less efficiency in probing for gas species on TRAPPIST-1 planets (T. J. Fauchez et al. 2019).

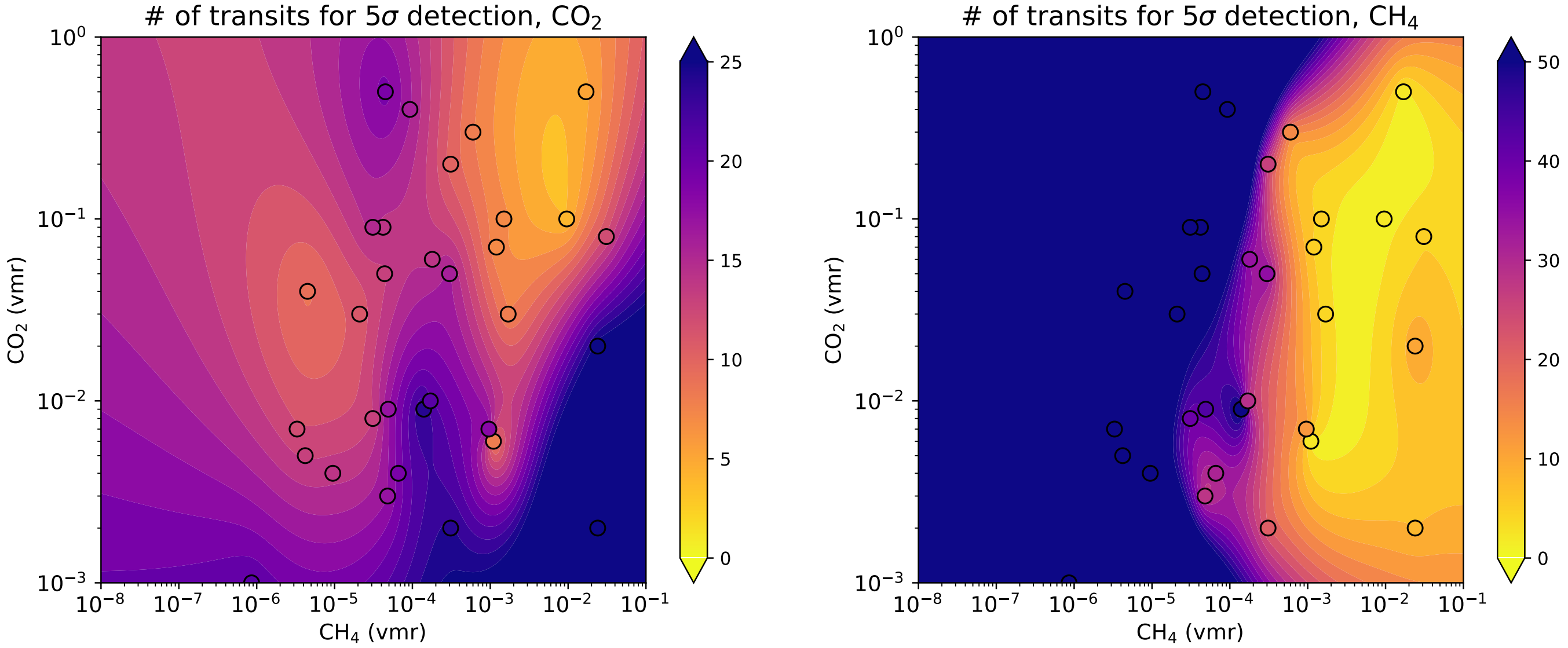

**Figure 15.** Kriging interpolation of the number of transits needed to detect $CO_2$ and $CH_4$ at $5\sigma$.

In some regions of our parameter space, positive detections of $CO_2$ and $CH_4$ are feasible with 10 or fewer transit observations. $CO_2$ detectability has two favored lobes, one at high $CO_2$ ($\geqslant 10^{-1}$) and high $CH_4$ ($\sim 10^{-1}$), the other at moderate $CO_2$ ($10^{-2} \leqslant CO_2 \leqslant 10^{-1}$) and low $CH_4$ ($\sim 10^{-5}$). Methane detectability uniformly requires high $CH_4$ ($\geqslant 10^{-3}$). The regions of detectability for both gases generally coincide with cold to cool climates (Figure 13). Cool and cold simulations have less water vapor and importantly fewer water clouds in their atmosphere (Figure 14), permitting larger amplitudes of transit signals to be detected (Figure 10). For low-$CH_4$ scenarios, a sweet spot exists for $CO_2$, where increasing $CO_2$ abundances makes its transmission signal stronger, but too much $CO_2$ can lead to a warm planet with increased cloudiness, which obscures transmission spectra. When high $CH_4$ is present, its strong antigreenhouse effect can suppress planet temperatures, keeping skies relatively free of water clouds and improving detectability prospects despite large $CO_2$ burdens. As long as hazes are not prevalent (e.g., $CH_4/CO_2 \leqslant 0.2$), a scenario with high $CO_2$ and high $CH_4$ together is the most amenable for detections of both gases.

Warm habitable climates are more difficult to characterize owing to their increased water vapor and water clouds. Similar conclusions have been drawn elsewhere about the role of water clouds in muting transmission signals (T. J. Fauchez et al. 2019; T. D. Komacek et al. 2020; G. Suissa et al. 2020a); however, here our multimodel sparse gridding approach allows us to visualize the broader picture and correlate between climate, clouds, and observables. $CO_2$ clouds do not form in our simulations, as atmospheric temperatures are universally above the $CO_2$ condensation point, even in our completely glaciated cases. However, $CO_2$ clouds and condensation could occur on TRAPPIST-1 e if it had a total atmospheric pressure much less than 1 bar and could occur on the outer TRAPPIST-1 planets for a wide range of atmospheric compositions (M. Turbet et al. 2018; T. J. Fauchez et al. 2019). Hazes are found to play a moderate role in modulating detectability thresholds. Hazes only exist on the far right of the panels ($CH_4 \geqslant 10^{-2}$). Hazes are found to considerably truncate $CO_2$ detectability in this region, particularly for low $CO_2$ amounts, increasing the number of transits required for a 5$\sigma$ detection by a factor of 2–4. Hazes also reduce $CH_4$ detectability by a factor of 5–10. However, the high $CH_4$ amounts needed to produce the haze in the first place permit its observation within fewer than 10 transits regardless.

We also notice a region of sharp sensitivity in detectability for each gas when $CO_2 \sim 10^{-2}$ and $10^{-4} \leqslant CH_4 \leqslant 10^{-3}$. Here features in Figure 15 do not correlate with any obvious variations in global mean climatological properties shown in Figures 13 and 14. A deeper exploration of model outputs reveals that significant variations are present in the 3D spatial structure of ice clouds at and above 1 mbar pressure levels. Furthermore, high-altitude variations in clouds are not well correlated with column-integrated amounts since most of the ice cloud mass resides below 10 mb. Recent work by T. J. Fauchez et al. (2025) argues that adequately capturing the spatial variations in high clouds is very important for ensuring proper modeling of continuum levels in transmission spectra and can strongly affect the detectability of gas features. Similar findings were also found when analyzing transmission spectra from the THAI project (T. J. Fauchez et al. 2022). We leave it for future studies to explore deeper into the nonlinear behaviors of high-altitude clouds and their outsized effects on transmission spectra.

In total, our results show that for TRAPPIST-1 e the climates states that would be most amenable to observations are cold and icy planets, with little water vapor, clouds, or haze present that would obscure transmission spectra. While hazes make it more difficult to observe the underlying gases, high-$CH_4$ contents should still be detectable within a reasonable number of transits even with the presence of haze included. Titan-like planets (i.e., frozen surface, high $CH_4$, hazy) would be amenable for positive $CH_4$ detections, coupled with analysis of any continuum slope at shortwave lengths that could help discern the presence of hazes (Figure 10). More broadly, visualization of various atmospheric properties and observational properties across a wide parameter space helps us better understand the terrain we are attempting to navigate, in connecting scant observations to climate states.

## 5. Discussions

### 5.1. Prospects for Atmospheric Retention

There is an ongoing debate as to whether the TRAPPIST-1 planets, or any terrestrial M dwarf planets for that matter, can retain their atmospheres in the face of the aggressive pre-main-sequence phase of low-mass stars and the possibility for sustained stellar activity throughout their lifetimes. A prolonged runaway greenhouse climate triggered by the high irradiances of the pre-main-sequence phase of M dwarfs could cause all the primordial water of a planet to be lost to space, leading to desiccated, abiotic, $O_2$-dominated atmospheres (R. Luger & R. Barnes 2015). Even after the pre-main-sequence phase has passed, continuous flaring from active M dwarf stars could erode away the atmosphere, rendering a planet airless (V. S. Airapetian et al. 2017). Recent observations with JWST point toward TRAPPIST-1 b likely being a bare rock (T. P. Greene et al. 2023), while thick $CO_2$-dominated atmospheres have been ruled out for TRAPPIST-1 c (S. Zieba et al. 2023).

However, J. Krissansen-Totton (2023) argues that the nondetection of atmospheres around the inner two TRAPPIST planets (b and c), both of which are significantly inward of the runaway greenhouse limit, does not broadly doom all of the TRAPPIST-1 planets to being barren airless rocks in space. The outer TRAPPIST planets could plausibly retain $CO_2$–$O_2$-dominated atmospheres (J. Krissansen-Totton & J. J. Fortney 2022). Prior theoretical work by C. Dong et al. (2018) argues similarly that the outer TRAPPIST planets (g and h specifically) could retain their atmospheres on billion-year timescales. The probability of the middle TRAPPIST planets (d, e, and f) to retain their atmospheres is likely sensitive to the specific rates of atmospheric loss processes and the initial gas inventories of each planet. For instance, E. Bolmont et al. (2017) show that differing assumptions on the extreme-UV (0.1–100 nm) emission can lead to an order-of-magnitude spread in the expected water loss from planets around M dwarfs. R. Luger et al. (2015) suggest an alternative pathway where a primordial mini-Neptune-like atmosphere could lose light gases, while leaving a remnant high mean molecular weight atmosphere bound to the solid planetary core. Even if the primordial atmospheres are stripped away completely, M dwarf terrestrial planets could source a secondary atmosphere through volcanic activity, perhaps driven by enhanced tidal interactions with their close host stars (E. S. Kite & M. N. Barnett 2020). Encouragingly, the JWST Director's Discretionary Time program has promised 500 hr of observing time for secondary eclipse and transit observations of rocky exoplanets, which will shed light on the possibility of atmospheres on M dwarf rocky planets.

### 5.2. Caveats on Ocean Treatment

When interpreting the results from our 3D climate model, `ExoCAM`, it is important to remember that we used a thermodynamic slab ocean with no transport and no continents. Other studies have shown that for temperate tidally locked planets ocean heat transport on a globally ocean-covered world could homogenize surface temperatures and increase the fraction of habitable surface area (Y. Hu & J. Yang 2014; A. D. Del Genio et al. 2019). Still, in other scenarios, sea-ice rafting by ocean currents could have the opposite effect and contribute to increased ice cover for cold tidally locked worlds (J. Yang et al. 2020). To further complicate matters, the presence of continents strongly modulates the nature of ocean heat transport, and in some cases it can negate the abovementioned effects entirely by preventing longitudinal transport. In simulations with Earth-like continents (A. D. Del Genio et al. 2019), an equatorial supercontinent (A. M. Salazar et al. 2020), or longitudinally extended continents (J. Yang et al. 2019), substellar-to-antistellar ocean heat transport is restricted and yields climate states that closely mirror results from static thermodynamic slab oceans. Work by J. Leconte (2018) considering tidal torques between star and planet argues that if emergent land exists on M dwarf planets, that land likely resides at substellar and antistellar points, similar to the boundary conditions assumed by A. M. Salazar et al. (2020). Given the significant uncertainties introduced by continents, the order-of-magnitude increase in computational time required to equilibrate a full dynamic ocean model, and the broader goals of our paper, we have opted to use a simple and fast representation of the ocean.

### 5.3. Considerations on Hazes

Several other recent works have been published on modeling hazes in exoplanetary atmospheres. M. T. Mak et al. (2024) modeled photochemical hazes on TRAPPIST-1 e using a combination of `Atmos` and the `UM` model developed by the Met Office. The primary differences in technical approach are that M. T. Mak et al. (2024) prescribed a horizontally uniform haze field in their 3D model based on results from 1D `Atmos` calculations, while we input zenith-angle-dependent haze production rates from `Atmos`, which are then acted on by `ExoCAM` dynamics and radiation, and `CARMA` microphysics, resulting in spatial nonuniformity of the haze as shown in Figure 9. Second, they assume Mie particles in all simulations as opposed to our base assumption, which assumes fractal aggregate particles. Despite these technical differences, a closer examination reveals important similarities. Both our work and M. T. Mak et al. (2024) find a strong stratospheric inversion driven by $CH_4$, with stratospheric temperatures of ∼260 K for high $CH_4$ conditions even when hazes are absent from the models. M. T. Mak et al. (2024) find a much more significant surface cooling effect attributed to hazes, with global mean surface cooling from thick hazes of ∼17.6 K, contrasting our ∼2 K surface cooling described in Section 3.3. However, if we compare apples to apples (i.e., Mie particles), are results converge. We find the global mean temperature for case 23 with no haze, fractal haze, and Mie haze to be 228.35, 226.08, and 213.71 K, respectively. Thus, the difference in radiative effect between fractal and Mie hazes is significant and likely explains the differences in our results. While we find slightly less cooling than M. T. Mak et al. (2024) even with Mie hazes, the residual difference may be accounted for by the horizontal distribution noted in Figure 9. Further analysis of our ensemble reveals that we also find that multiple atmospheric circulation regimes are present dependent on the atmospheric composition, in agreement with M. T. Mak et al. (2024) and as first noted by D. E. Sergeev et al. (2022b). Here we find three primary emergent dynamical regimes: a single equatorial jet state, a double midlatitude jet state with counterrotational flow aloft, and a triple jet state with two midlatitude jets in the troposphere and strong equatorial jet aloft. Future work is needed to dive deeper into the reasons for the multiple equilibria emerging in the atmospheric circulation of TRAPPIST-1 e and for comparing divergent results among various exoplanetary general circulation models.

In another recent study, M. Cohen et al. (2024) used the `ExoPlaSim` model with radiatively active advected tracers to examine how dynamics would change the global haze distribution. They find a high sensitivity between the atmospheric circulation regime of planets at varying rotation periods, planetary radii, and gravity and the location of haze mass accumulation. In comparing like simulations (i.e., TRAPPIST-1 e with a 6-day rotation period; Figure 8(a) in M. Cohen et al. 2024), qualitative agreement is found in the haze column being greatest at high latitudes and in gyres in the eastern hemisphere. Lastly, M. G. Lodge et al. (2024) performed a comprehensive modeling study examining assumptions involving the optical properties of fractal aggregate aerosols. They concluded that the choice of refractive indices, the specific fractal aggregate proprieties (monomer size, fractal dimension, etc.), and the specific modeling tool used for computing the extinction and scattering for aggregates based on these assumptions all have a significant effect on the radiative properties of fractal aerosols. Future work at the intersection of exoplanet observations, laboratory haze experiments, and theoretical modeling of hazes should be a fruitful path for interested researchers.

### 5.4. Implications for Biosignature Detection and Interpretation

Molecular $O_2$ is likely not detectable on TRAPPIST-1 e at Earth-like abundances with JWST, due to the size and strength of its strongest feature at 0.76 $\mu$m, JWST's limitations in sensitivity and spectral resolution at this wavelength, and poor backlighting from the star short of its near-IR peak emission (J. Lustig-Yaeger et al. 2019). Similarly, detecting $O_3$ at Earth-like abundances will be challenging (though perhaps not impossible), due to the predicted strength of its 4.8 and 9.65 $\mu$m band features and poor SNRs at mid-IR wavelengths from the decline of the star's thermal emission beyond ∼2 $\mu$m (J. Lustig-Yaeger et al. 2019). Therefore, the most promising indicator of a global biosphere on TRAPPIST-1 e could be the simultaneous presence of $CH_4$ and $CO_2$ with low CO (J. Krissansen-Totton et al. 2018; T. Mikal-Evans 2021; M. A. Thompson et al. 2022; Y. Watanabe & K. Ozaki 2024), which could indicate a methanogenic biosphere and could apply even in $O_2$-rich cases, where the $O_2$ is effectively spectrally invisible (V. S. Meadows et al. 2023). A detectability advantage for atmospheric $CH_4$ is conferred by the UV spectrum of M dwarf stars, which allow for greater photochemical accumulation at a given production rate versus that expected for habitable planets orbiting Sun-like host stars, which have stronger near-UV emission owing to their hotter photospheric temperatures. This insight was originally demonstrated by A. Segura et al. (2005) and reproduced by many subsequent researchers (e.g., S. Rugheimer et al. 2015; V. S. Meadows et al. 2018; F. Wunderlich et al. 2019), though we note that the photochemical mechanisms for enhanced $CH_4$ accumulation, as well as its predicted magnitude, vary between $O_2$-rich and anoxic atmospheric scenarios (A. Akahori et al. 2024; W. Broussard et al. 2024).

Here we modeled anoxic, Archean-like atmospheric scenarios for TRAPPIST-1 e, treating both $CO_2$ mixing ratios and $CH_4$ production fluxes as independent variables. One of the core findings of our study is that the antigreenhouse effect of $CH_4$ can lead to significant global cooling, even at high $CO_2$ abundances. Due to the reduction in habitable (open ocean) area, it is likely that negative feedbacks would reduce biological methane production at $CH_4$ mixing ratios of ∼0.1%–1% or more, depending on $CO_2$ mixing ratio and other factors. This corresponds roughly to $CH_4$ production rates comparable in magnitude to those of modern Earth, ∼30–40 Tmol yr$^{-1}$ (E. J. Dlugokencky et al. 2011). Such negative feedbacks would likely apply even for planets that maintain some fractional habitability near the substellar point, because of the overall reduction in energy available to power a photosynthetic biosphere and produce biomass. Correspondingly, this may reduce the probability that biogenic $CH_4$ accumulates beyond several hundred parts per million on TRAPPIST-1 e, or similar planets orbiting ultracool dwarfs. While this extrapolation is speculative (as is any biosignature scenario on an exoplanet), it calls into question the usual assumptions used to predict possible biosignature gas abundances, namely that the integrated and averaged surface production flux of Earth today (or in the past) is a reasonable starting assumption. The possible reduction of fractional surface habitability due to the $CH_4$ antigreenhouse may similarly inhibit the production and accumulation of other biosignature gases, possibly leading to the potential for false negatives for biosignature detection (C. T. Reinhard et al. 2017). Moreover, the deposition of photochemical by-products such as CO and their possible consumption by life could be inhibited by substantial global ice cover. Future work that directly couples metabolic and climate models may afford further insight into this possibility. We emphasize that the antigreenhouse effect of $CH_4$ found here for planets orbiting ultracool dwarfs may have substantial implications for detecting and interpreting potential atmospheric biosignatures with transmission spectroscopy.

### *5.5. Perspectives on Using Climate Models to Interpret Observations*

Exoplanet discoveries have now become routine in astronomy, and each new discovery raises questions about the characteristics of any planetary atmosphere that might exist. Observational data available for newly discovered exoplanets can include planetary mass, radius, orbital distance, and spectral type of the host star; these variables are a necessary but insufficient condition for defining the atmosphere of a planet using a climate model. Spectral observations will ultimately be required to place any meaningful constraints on the atmospheric composition and properties of an exoplanet. But such spectral observations will only occur for a limited number of salient targets, and astronomers who discover an exoplanet today may wish to explore its potential characteristics or habitability without needing to wait for future observing facilities to be developed.

The desire to make even tentative statements about the climates of newly discovered planets has motivated numerous studies that consider ad hoc combinations of climate states as possibilities (e.g., M. Turbet et al. 2016; E. T. Wolf 2017; G. Suissa et al. 2020b; L. Biasiotti et al. 2024). For example, the study by G. Suissa et al. (2020b) used `ExoCAM` to simulate 20 different climate states for the habitable zone planet TOI-700 d: this included modern Earth, Archean Earth, and early Mars configurations at different pressures and with both aquaplanet and desiccated conditions, as well as a hydrogen-supporting case and a generic case with nitrogen only. This range of possibilities can demonstrate different spectral features that theoretically could be used to distinguish between these climate states, but the requisite high-precision observations will not be possible with any near-term facilities. Another example involved the candidate planet Gliese 581 g, which was later refuted by further analysis: climate modeling studies of this now nonexistent planet were conducted during the interim between discovery and follow-up observations, in an effort to suggest the possibility that this planet could be habitable (e.g., R. T. Pierrehumbert 2010; P. von Paris et al. 2010; K. Heng & S. S. Vogt 2011). Modeling individual exoplanet atmospheres without knowledge of detailed planetary spectra can help to promote awareness of particular planets of interest, but such an activity may have diminishing scientific returns.

In most cases, such results do not provide any novel insights beyond the existing calculations of the generalized circumstellar habitable zone (e.g., J. F. Kasting et al. 1993; R. K. Kopparapu et al. 2013, 2014; R. Kopparapu et al. 2017) that can already be leveraged to make statements about habitability for new discoveries. These prior calculations explored a parameter space spanning stellar spectral type, orbital distance, planetary mass, and atmospheric pressure with varying abundances of carbon dioxide. The limitation in previous calculations is the extent to which other variables have been explored, such as additional greenhouse gases and the effects of topography, or the consequences of initial conditions (e.g., M. Turbet et al. (2021, 2023). Systematic exploration of any of these properties will generally be more scientifically useful than single-planet studies, as they can be relevant toward interpreting a broad range of observations.

The ultimate goal of the multimodel pipeline demonstrated in this study is to develop a community data product that interpolates habitability limits and atmospheric properties across a broad multidimensional parameter space of interest to the observation of M dwarf planets. We have shown the feasibility in this study of interpolating sparse samples across a two-dimensional parameter space. Such methods can be extended to an arbitrary number of dimensions as needed. Once complete, this approach will alleviate the need for conducting single-planet climate simulations, as any new exoplanet discovery will be able to draw upon the multidimensional interpolated parameter space to infer possible atmospheric properties without requiring any new simulations.

## 6. Conclusions

In this study we have performed a multimodel ensemble to explore the chemistry, climate, and transmission spectra of TRAPPIST-1 e. First, we used a 1D photochemical model (`Atmos`) to compute a 480-simulation grid of self-consistent $N_2$, $CO_2$, $CH_4$, and haze atmospheric compositions. Then, we used a QMC approach to formulate a 32-simulation sparse sampling from the original 480 atmospheric compositions to explore with our computationally expensive 3D climate model (`ExoCAM`). Cases with $CH_4/CO_2 \geqslant 0.2$ were shown by `Atmos` to have optically significant photochemical hazes. For these cases, `ExoCAM` was fully coupled to a size-bin-resolved aerosol microphysical model (`CARMA`) to simulate fractal aggregate hazes, with mass production rates from `Atmos`. Results from our 32 climate simulations, including profiles of temperature, gas mixing ratios, clouds, and hazes, were then used to compute transmission spectra using the `Planetary Spectrum Generator`. After computing results for climate and transmission spectra for our 32-simulation sparse sampling, these results were interpolated across the broader grid using a technique known as kriging, which is closely related to Gaussian process emulation. Kriging interpolation yielded high-resolution maps of important climate and observable properties across our parameter space, allowing us to make correlations between achievable observations and predictions for the underlying atmospheric state. Our use of a pipeline of well-established publicly available models lends confidence to the self-consistency of our solutions. Our use of a sparse grid to guide 3D climate simulations saved an order of magnitude in computational and real-world costs. Finally, kriging interpolation allowed us to smoothly fill in all spaces of our grid.

TRAPPIST-1 e was the most logical choice for an object of study, as it is a prime habitable zone terrestrial exoplanet target and will be one of few that is amenable to characterization in the near term. Our modeling results revealed several interesting features that are connectable to observations. As expected,

TRAPPIST-1 e is durably habitable and can maintain some amount of liquid surface water for a wide variety of $CO_2$, $CH_4$ combinations. Increasing $CO_2$ universally warms the planet surface. However, the effects of $CH_4$ are more complicated. $CH_4$ can contribute additional warming via its greenhouse effect for mixing ratios up to $\sim10^{-3}$. However, at larger abundances strong near-infrared absorption in the stratosphere by $CH_4$ triggers an antigreenhouse effect, cooling the planet surface while forming a broad upper atmosphere inversion. This phenomenon is accentuated by the red spectrum of TRAPPIST-1, which peaks near 1 $\mu$m. When $CH_4/CO_2 \geqslant 0.2$, photochemical hazes would become important for TRAPPIST-1 e. However, in this study we found photochemical hazes to have a limited effect on the overall climate owing to a combination of factors. Despite haze production occurring only on the sunlight hemisphere, atmospheric circulation causes the haze mass to accumulate predominantly at the poles and the nightside of the planet, where stellar irradiances are small and nonexistent. Furthermore, we found that our particular choices for refractive indices and the fractal aerosol assumptions cause the haze to have a deep minimum in absorption aligning precisely with where TRAPPIST-1 emits the most energy. Thus, both atmospheric circulation properties and aerosol optical properties limit the impact of the haze on incoming stellar radiation. Future studies on exoplanetary hazes would benefit from exploring the many varieties of assumptions possible for considering haze optical properties and their potential for unique interactions with stellar radiation and atmospheric circulation.

For our high-$CH_4$ simulations ($CH_4 \geqslant 10^{-3}$), which are prone to a strong antigreenhouse effect, we show that interpreting disk-averaged thermal emission and albedo quantities can lead to misleading conclusions about the surface climate. Here our high-$CH_4$ simulations tend to have the lowest albedos and highest thermal emission despite having the coldest surface temperatures, the opposite of common intuitions. Here photochemical hazes do not play a significant role in modifying the planetary radiation fields, but for other refractive index and stellar energy distribution combinations, hazes could contribute more strongly to the antigreenhouse effect and albedo and thermal emission modifications. One should be cautious with respect to assumptions made regarding exoplanet surface climate based on zeroth-order energy balance measurements.

$CO_2$ and $CH_4$ could be feasibly detected on TRAPPIST-1 e with JWST with 5$\sigma$ confidence in $\sim$10 transits, but only under certain circumstances. In less ideal circumstances, 5$\sigma$ detections of $CO_2$ and $CH_4$ could take dozens, or even hundreds, of transits. $H_2O$ remains too difficult to detect in all scenarios, requiring hundreds to thousands of transits to detect regardless of circumstance. Transmission spectral results calculated here show that colder climate states are easier to characterize than warmer states, due to being water limited and having few water clouds. The paucity of water clouds permits transmission spectra to probe down to lower layers in the atmospheres. However, our coldest climates also form hazes, which counteract the absence of clouds and decrease detectability. All things considered, $CO_2$ has two regions of favored detectability in our parameter space: one at high $CO_2$ ($\geqslant10^{-1}$) and high $CH_4$ ($\sim10^{-1}$), and the other at moderate $CO_2$ ($10^{-2} \geqslant CO_2 \leqslant 10^{-1}$) and low $CH_4$ ($\sim10^{-5}$). Methane detectability uniformly requires high $CH_4$ ($\geqslant10^{-3}$). The detectability of $CO_2$ and $CH_4$ depends on a delicate balance between their mixing ratios and line strengths and their control on clouds and hazes, through greenhouse and antigreenhouse forcings on climate and through photochemical haze formation.

## Acknowledgments

This material is based on work performed as part of the Consortium on Habitability and Atmospheres of M dwarf Planets (CHAMPs) team, supported by the National Aeronautics and Space Administration (NASA) under grant Nos. 80NSSC21K0905 and 80NSSC23K1399 issued through the Interdisciplinary Consortia for Astrobiology Research (ICAR) program. E.T.W. additionally acknowledges support from the NASA Habitable Worlds program grants 80NSSC20K1421 and 80NSSC21K1718. E.W.S. and M.L. additionally acknowledge support from the NASA Exoplanet Research Program via grant Nos. 80NSSC22K0235 and 80NSSC23K0039. Some computations were performed using the computer clusters and data storage resources of the UCR-HPCC, which were funded by grants from NSF (MRI-2215705, MRI-1429826) and NIH (1S10OD016290-01A1). T.J.F., R.K., and G.L.V. acknowledge support from the GSFC Sellers Exoplanet Environments Collaboration (SEEC), which is supported by the NASA Planetary Science Division's Internal Scientist Funding Model. S.P. acknowledges support from NASA under award No. 80GSFC21M0002.

*Software:* `Atmos` (G. Arney et al. 2016), `ExoCAM` (E. T. Wolf et al. 2022), `CARMA` (R. P. Turco et al. 1979), `Planetary Spectrum Generator` (G. L. Villanueva et al. 2018), `GlobES` (T. J. Fauchez et al. 2025), PyKrige (B. Murphy et al. 2022), CESM (R. B. Neale et al. 2013).

## Author Contributions

It need be noted that this was truly a collective effort. While E.T.W. was the lead writer, he acknowledges the large fraction of work performed by coauthors. E.T.W. ran `ExoCAM` and `CARMA` simulations, performed climate analysis, and was the primary writer of this manuscript. E.W.S. ran `Atmos` simulations, performed photochemical analysis, and contributed to writing. J.H.-Q. formulated the QMC sparse sampling grid, performed kriging analysis, and contributed to writing. T.J.F ran `PSG` simulations, performed transmission spectral analysis, and contributed to writing. S.T.B. ran `Atmos` for hazy cases, produced haze production rate files necessary for `ExoCAM-CARMA` coupled simulations, and contributed to writing. M.L. contributed to running and analyzing `Atmos` simulations. G.V. modified `PSG` to accept user-specified fractal aggregate haze optical property files. S.P. contributed stellar spectra for TRAPPIST-1. R.K.K. is the science PI of CHAMPs, helped in devising the study, and was instrumental in convincing the lead author not to quit academia, although recent events mean that that option is back on the table.

## Appendix

The appendix contains ancillary tables and plots which are aid in interpreting the results presented in the main text. Table A1 describes the surface boundary conditions for photochemical modeling. Table A2 describes global mean error quantities from kriging interpolations using subsets of our sparse sampling points. Figure A1 shows the kriging interpolation for surface temperature as it evolves with inclusion of different sequences shown in Figure 2. Figures A2, A3, and A4 show the root off the kriging variances. Discussion of these figures occurs in Section 4 of the main text.

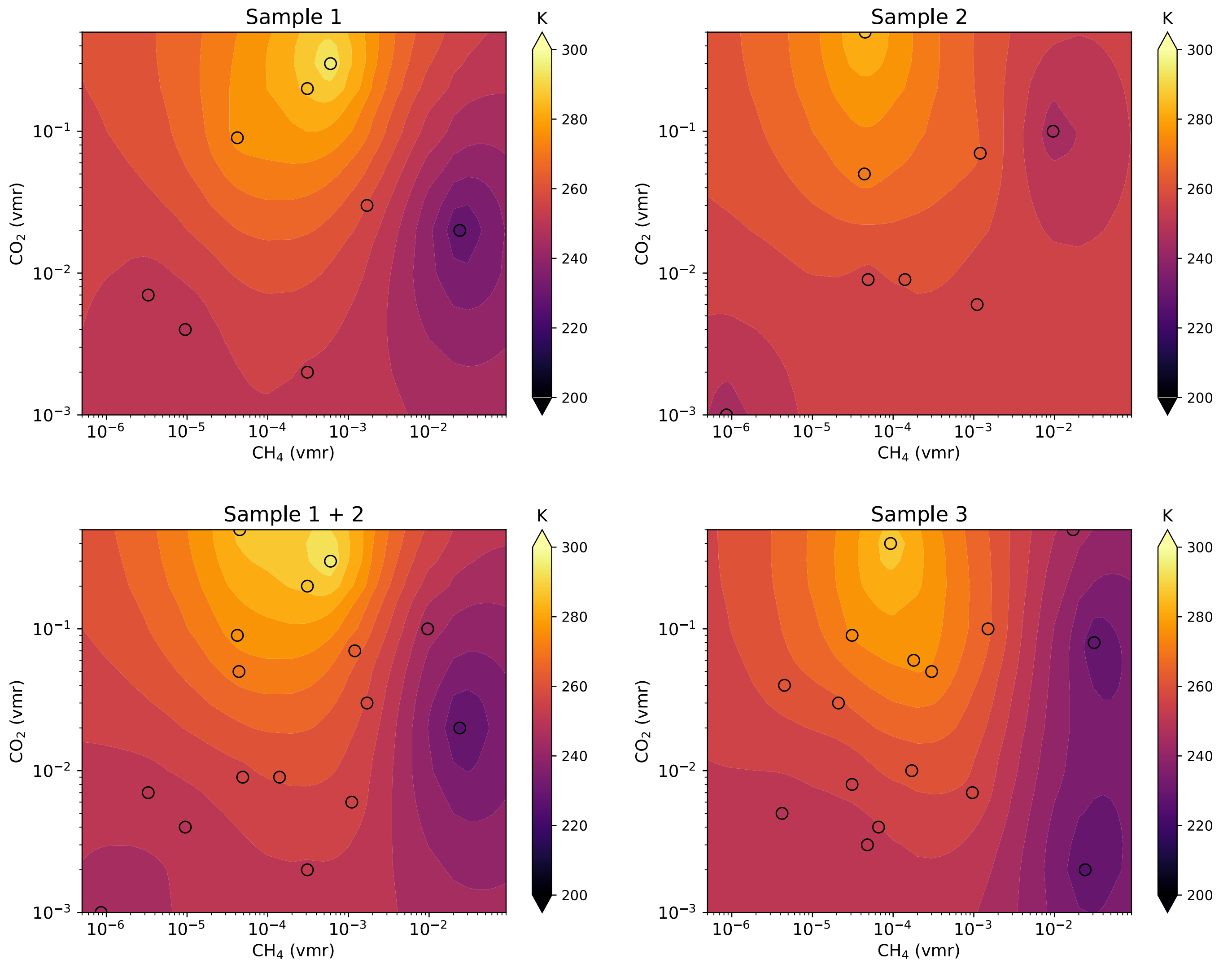

**Figure A1.** Kriging interpolation of surface temperature for the 32 `ExoCAM` simulations, with separate interpolations shown for Sample 1, Sample 2, Sample 1+2, and Sample 3. The black circles show the `ExoCAM` simulations. With each additional sequence the emulation of the parameter space is refined.

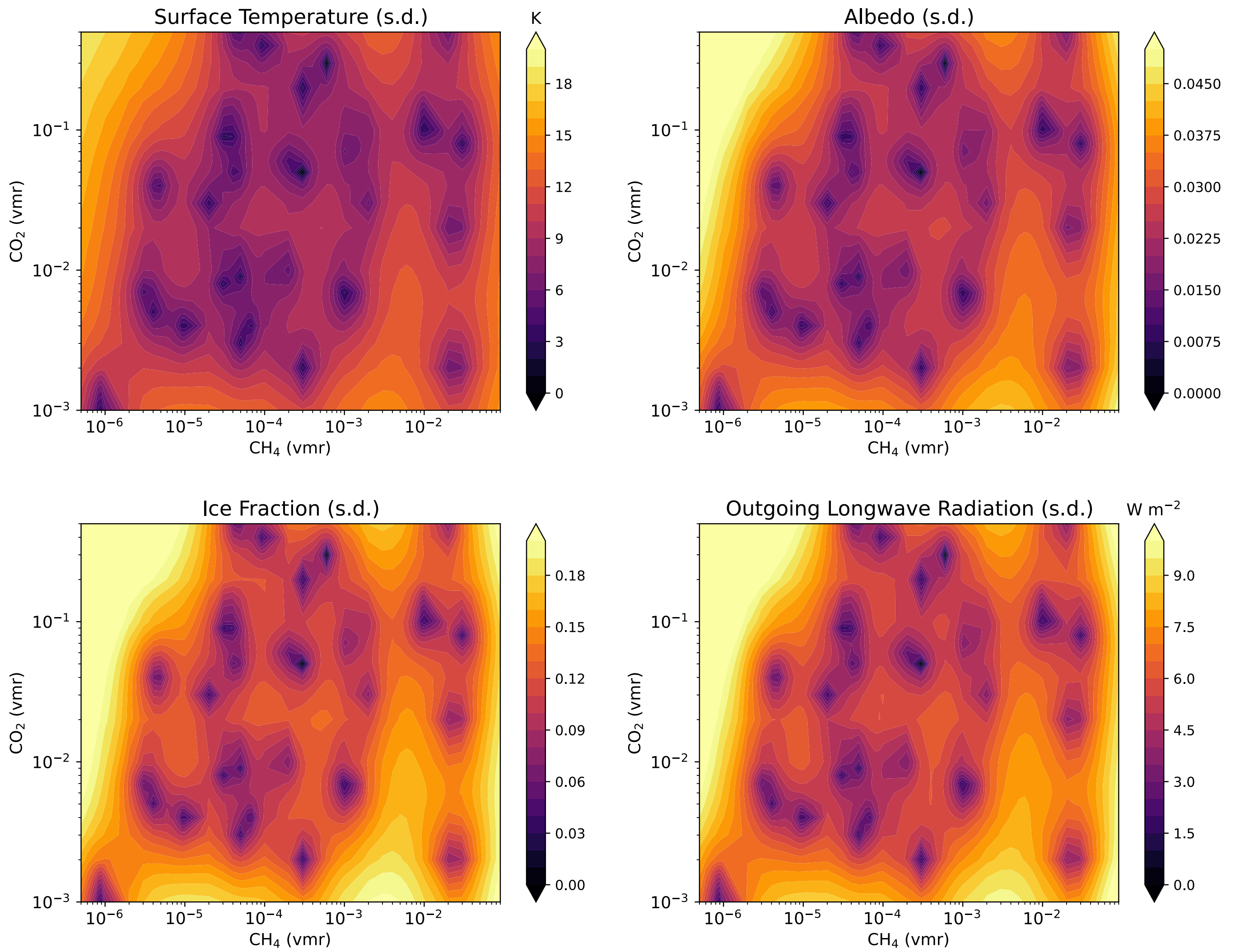

**Figure A2.** Root of the kriging variance across the parameter space from the kriging interpolation of the 32 `ExoCAM` simulations. Panels are shown and labeled for the global mean surface temperature, albedo, ice fraction, and OLR.

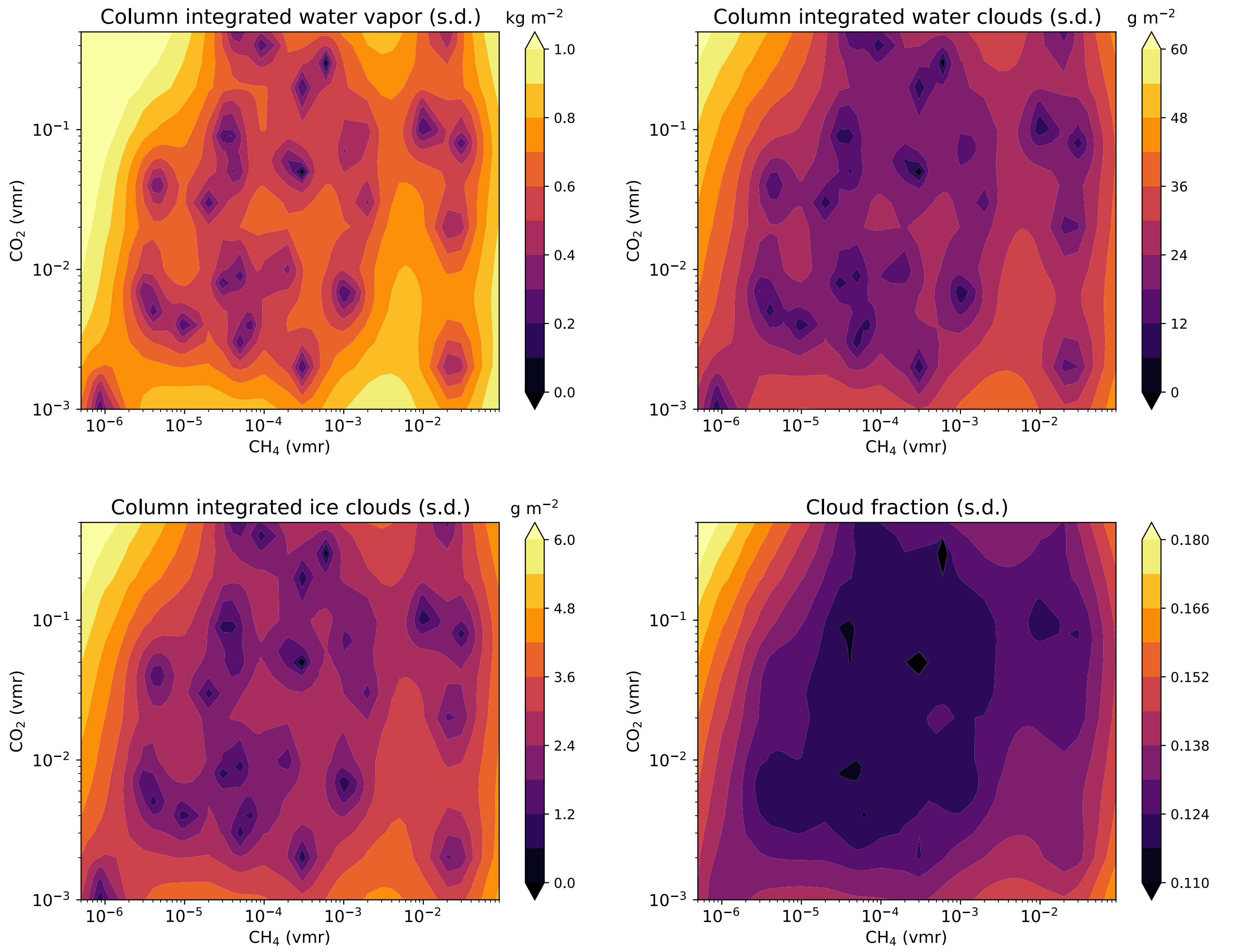

**Figure A3.** Root of the kriging variance across the parameter space from the kriging interpolation of the 32 ExoCAM simulations. Panels are shown and labeled for the column-integrated water vapor, liquid water clouds, ice water clouds, and cloud fraction.

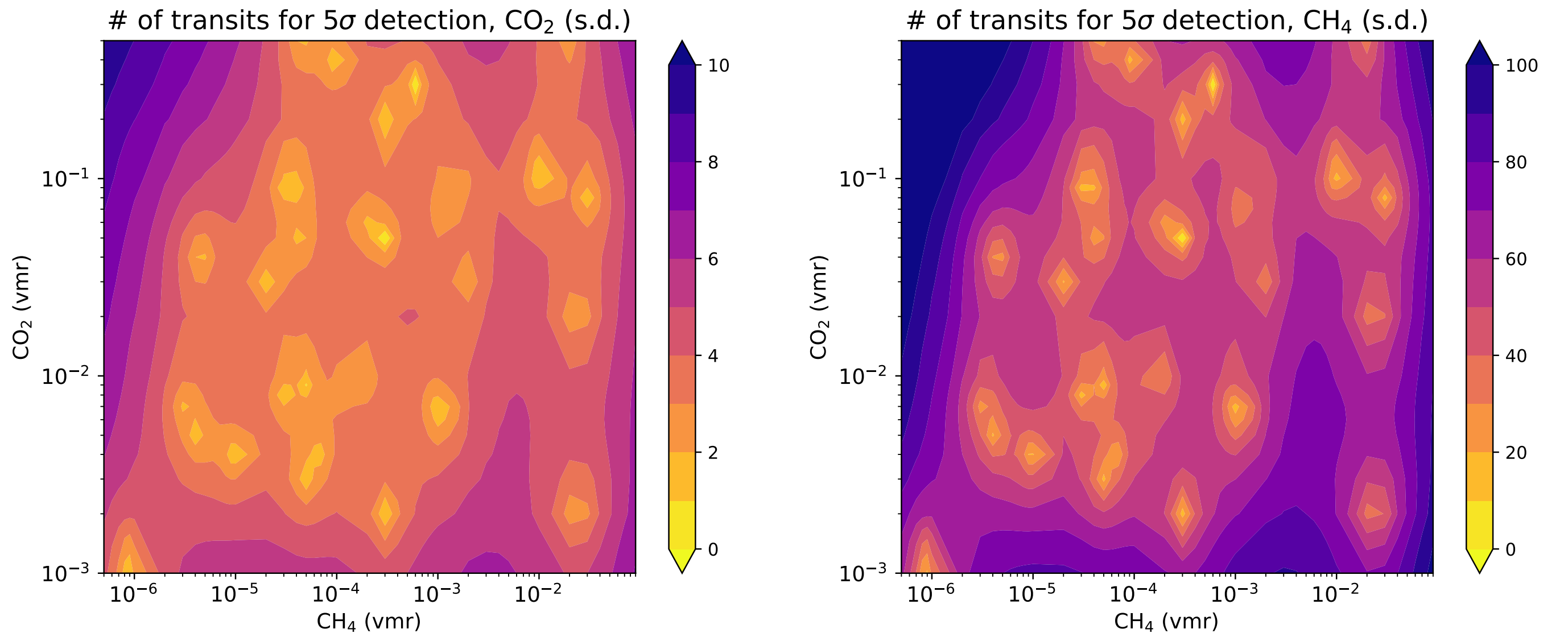

**Figure A4.** Root of the kriging variance across the parameter space from the kriging interpolation of the number of transits needed to detect $CO_2$ and $CH_4$ at $5\sigma$.

**Table A1**
Surface Boundary Conditions for the Atmos Photochemical Modeling Described in Sections 2.1 and 3.1

| Chemical Species[a] | Deposition Velocity (cm $s^{-1}$) | Flux (molecules $cm^{-2}$ $s^{-1}$) | Mixing Ratio |
|---|---|---|---|
| O | 1.0 | ⋯ | ⋯ |
| $O_2$ | ⋯ | ⋯ | ⋯ |
| $N_2$ | ⋯ | ⋯ | fill gas |
| $CO_2$ | ⋯ | ⋯ | variable |
| $H_2O$ | ⋯ | ⋯ | fixed[b] |
| H | 1.0 | ⋯ | ⋯ |
| OH | 1.0 | ⋯ | ⋯ |
| $HO_2$ | 1.0 | ⋯ | ⋯ |
| $H_2O_2$ | 0.2 | ⋯ | ⋯ |
| $H_2$ | $2.4 \times 10^{-4}$ | $3.0 \times 10^{10}$ | ⋯ |
| CO | $1.2 \times 10^{-4}$ | ⋯ | ⋯ |
| HCO | 1.0 | ⋯ | ⋯ |
| $H_2CO$ | 0.2 | ⋯ | ⋯ |
| $CH_4$ | 0.0 | variable | ⋯ |
| $CH_3$ | 1.0 | ⋯ | ⋯ |
| $C_3H_3$ | 0.1 | ⋯ | ⋯ |
| NO | $3.0 \times 10^{-4}$ | $1 \times 10^{9}$ | ⋯ |
| $NO_2$ | $3.0 \times 10^{-3}$ | ⋯ | ⋯ |
| HNO | 1.0 | ⋯ | ⋯ |
| $H_2S$ | 0.02 | $3.5 \times 10^{8}$ | ⋯ |
| $SO_2$ | 1.0 | $3.5 \times 10^{9}$ | ⋯ |
| $H_2SO_4$ | 1.0 | ⋯ | ⋯ |
| HSO | 1.0 | ⋯ | ⋯ |
| $O_3$ | 0.07 | ⋯ | ⋯ |
| $HNO_3$ | 0.2 | ⋯ | ⋯ |
| $SO_4$AER | 0.01 | ⋯ | ⋯ |
| $S_8$AER | 0.01 | ⋯ | ⋯ |
| HCAER | 0.01 | ⋯ | ⋯ |
| HCAER2 | 0.01 | ⋯ | ⋯ |

**Notes.**
[a] Species included in the photochemical scheme with a deposition velocity and flux of 0 include $C_2H_6$, N, $CH_3H_2$, $CH_3C_2H$, $CH_2CCH_2$, $C_3H_5$, $C_2H_5CHO$, $C_3H_6$, $C_3H_7$, $C_3H_8$, $C_2H_4OH$, $C_2H_2OH$, $C_2H_5$, $C_2H_4$, CH, $CH_3O2$, $CH_3O$, $CH_2CO$, $CH_3CO$, $CH_3CHO$, $C_2H_2$, $CH_2$, $HC_2$, $CS_2$, CS, OCS, S, HS, $SO_3$, SO, and $S_2$, $S_3$, and $S_4$.
[b] The tropospheric $H_2O$ profile is fixed assuming a 275 K surface with a relative humidity of 80%.

**Table A2**
Percent Error for Global Mean Quantities from the Using Kriging Interpolations Based on Sample 1 (Sample 1+2) When Compared with the Actual Values from the Simulations in Samples 2 and 3 (Sample 3)

| | Cases | $T_S$ | Albedo | Ice Fraction | OLR | Water Vapor | Water Clouds | Ice Clouds | Cloud Fraction | $CO_2$ Transits | $CH_4$ Transits |
|---|---|---|---|---|---|---|---|---|---|---|---|
| Sample 2 | 9 | −3.55% | 3.90% | 13.3% | −1.42% | −75.4% | −19.3% | 7.64% | −2.27% | 30.9% | 79.1% |
| See Sample 1 | 10 | −5.21% | −1.27% | −76.1% | 3.50% | −151% | −261% | 32.5% | −119% | −262% | −75.5% |
| | 11 | 0.63% | −3.33% | 4.48% | −0.77% | 19.6% | 19.7% | 13.4% | 11.6% | 39.5% | 30.6% |
| | 12 | 0.21% | −1.73% | 3.44% | −0.43% | −11.9% | −0.22% | 8.58% | −5.58% | −11.4% | −20.6% |
| | 13 | −0.57% | −3.81% | 3.65% | −0.92% | 0.02% | 6.90% | 4.90% | 0.85% | 14.7% | −42.1% |
| | 14 | −1.49% | 3.07% | 9.78% | −1.09% | 10.9% | −13.2% | −4.36% | 6.57% | −107% | −186% |
| | 15 | 4.12% | −1.17% | −4.34% | 0.54% | 58.4% | 43.5% | 19.33% | 22.9% | −81.6% | −628% |
| | 16 | 5.81% | −1.01% | −15.8% | 1.32% | 57.7% | −13.1% | −27.0% | 14.7% | 23.8% | 73.7% |
| Sample 3 | 17 | −0.52% | 0.36% | 1.12% | −0.44% | −26.1% | −7.25% | 3.05% | −8.18% | −11.4% | 15.2% |
| See Sample 1 | 18 | −8.58% | −34.6% | −90.3% | 2.67% | −345% | −248% | 43.9% | −102% | −190% | −165% |
| | 19 | 1.10% | 2.00% | 2.75% | −0.28% | 25.7% | 22.2% | 15.2% | 12.9% | 30.9% | −21.2% |
| | 20 | 0.88% | 4.14% | 6.00% | −0.73% | −2.26% | 9.57% | −17.3% | 13.2% | −3.50% | −37.2% |
| | 21 | −0.93% | 4.70% | 3.54% | −1.06% | −7.15% | 3.36% | 1.53% | −2.74% | −11.6% | −77.0% |
| | 22 | 1.27% | 0.03% | 2.52% | −0.001% | 22.5% | 23.0% | 8.16% | 15.1% | 9.41% | 10.2% |
| | 23[a] | −10.6% | −29.5% | −30.0% | 3.68% | −576% | −1630% | −90.1% | −150% | 70.4% | −109% |
| | 24 | 0.82% | 0.84% | 0.77% | −0.09% | 7.89% | −1.57% | 3.05% | −2.68% | 3.45% | 28.1% |
| | 25 | −0.65% | 5.66% | 4.35% | −0.91% | −6.97% | 0.57% | 13.4% | −4.83% | −31.7% | −54.9% |
| | 26[a] | −10.9% | −21.1% | −40.7% | 1.88% | −726% | −2090% | −5.43% | −225% | −20.9% | 41.3% |
| | 27 | −1.00% | 5.78% | 6.27% | −1.40% | −4.94% | 7.35% | 3.69% | 3.30% | 23.6% | −52.9% |
| | 28 | 4.72% | −14.8% | −20.1% | 1.57% | 58.4% | −13.7% | −27.6% | 15.5% | 9.52% | 48.2% |
| | 29 | −1.72% | 7.32% | 8.33% | −1.74% | −16.9% | 2.75% | 3.02% | 0.60% | 14.6% | −84.1% |
| | 30 | −2.24% | 10.5% | 9.08% | −1.13% | 2.47% | −25.7% | −2.66% | 0.40% | −107% | −159% |
| | 31 | 4.46% | −3.37% | −3.23% | 0.60% | 58.9% | 44.4% | 19.2% | 23.7% | 19.3% | −27.5% |
| | 32 | 0.26% | 3.65% | 6.73% | −0.51% | −8.16% | −7.53% | 11.2% | −11.5% | −60.9% | 18.7% |
| Sample 3 | 17 | −0.11% | −0.06% | −0.29% | −0.31% | −6.99% | −7.71% | 2.24% | −4.14% | −17.7% | −10.4% |
| See Sample 1+2 | 18 | −6.77% | 1.73% | −37.8% | 0.16% | −231% | −154% | 29.6% | −63.5% | −146% | −866% |
| | 19 | 0.44% | −2.16% | −2.07% | 0.49% | 4.89% | 9.98% | 2.88% | 4.07% | 27.6% | −60.4% |
| | 20 | 0.95% | −1.84% | −0.67% | 0.03% | −20.5% | 10.4% | −23.2% | 8.79% | 2.65% | 32.0% |
| | 21 | −0.49% | 1.55% | 0.94% | −0.37% | −2.43% | −2.08% | −2.39% | −3.67% | −17.5% | −26.6% |
| | 22 | 1.58% | −8.29% | −6.97% | 1.01% | 3.74% | 24.1% | 4.91% | 10.2% | 15.4% | 66.7% |
| | 23[a] | −11.9% | −25.9% | −25.6% | 3.23% | −748% | −1940% | −108% | −167% | 68.4% | −185% |
| | 24 | 0.10% | 1.66% | 2.39% | −0.25% | −3.79% | 0.67% | 4.82% | −5.74% | 5.95% | 4.55% |
| | 25 | −0.85% | 3.80% | 3.29% | −0.66% | 0.61% | −0.33% | 9.67% | −3.49% | −31.7% | −34.0% |
| | 26[a] | −6.42% | 13.4% | 10.1% | −1.14% | −254% | −1210% | −49.5% | −106% | −6.76% | 58.1% |
| | 27 | −0.84% | 2.82% | 2.16% | −0.69% | −10.2% | −0.83% | −3.80% | −3.67% | 17.1% | −160% |
| | 28 | 1.43% | −6.99% | −9.18% | 0.74% | 26.9% | −5.08% | −10.8% | 6.99% | 15.4% | −53.5% |
| | 29 | −1.27% | 4.63% | 3.96% | −1.03% | −15.1% | −3.27% | −3.13% | −5.36% | 6.63% | −271% |
| | 30 | −0.66% | 6.56% | 6.77% | −0.78% | −0.81% | −8.83% | −3.56% | 2.19% | −81.2% | 45.7% |
| | 31 | 1.19% | −1.18% | −1.04% | 0.30% | 19.1% | 21.9% | 3.28% | 14.5% | 17.6% | 59.9% |
| | 32 | −0.58% | 5.14% | 10.8% | −0.86% | −13.1% | −4.46% | 14.9% | −13.9% | −62.6% | −32.0% |

**Note.**
[a] 3D simulation contains photochemical hazes.

## ORCID iDs

Eric T. Wolf https://orcid.org/0000-0002-7188-1648
Edward W. Schwieterman https://orcid.org/0000-0002-2949-2163
Jacob Haqq-Misra https://orcid.org/0000-0003-4346-2611
Thomas J. Fauchez https://orcid.org/0000-0002-5967-9631
Sandra T. Bastelberger https://orcid.org/0000-0003-2052-3442
Michaela Leung https://orcid.org/0000-0003-1906-5093
Sarah Peacock https://orcid.org/0000-0002-1046-025X
Geronimo L. Villanueva https://orcid.org/0000-0002-2662-5776
Ravi K. Kopparapu https://orcid.org/0000-0002-5893-2471